\documentclass[astronomy,review,accept,pdftex,oneauthor]{Definitions/mdpi} 
\graphicspath{{./fig/}}
\firstpage{68} 
\pubvolume{3}
\issuenum{2}
\articlenumber{6}
\pubyear{2024}
\copyrightyear{2024}
\externaleditor{Academic Editor: Paolo Salucci}
\datereceived{31 January 2024} 
\daterevised{1 March 2024} 
\dateaccepted{1 April 2024} 
\datepublished{5 April 2024} 
\hreflink{https://\linebreak  doi.org/10.3390/astronomy3020006} 

\newcommand{\prl}{Phys. Rev. Lett.}

\newcommand{\araa}{Ann. Rev. Astron. Astrophys.}
\newcommand{\nat}{Nature}

\newcommand{\prd}{Phys. Rev. D}

\Title{Refracted Gravity Solutions from Small to Large Scales}

\TitleCitation{Refracted Gravity Solutions from Small to Large Scales}

\Author{Valentina
 Cesare \orcidA{}}

\AuthorNames{Valentina Cesare}

\AuthorCitation{Cesare, V.}

\address[1]{National Institute for Astrophysics, Astrophysical Observatory of Catania, Via Santa Sofia 78, \mbox{95123 Catania, CT, Italy;} valentina.cesare@inaf.it}

\corres{Correspondence: valentina.cesare@inaf.it}

\abstract{If visible matter alone is present in the Universe, general relativity (GR) and its Newtonian weak field limit (WFL) cannot explain several pieces of evidence, from the largest to the smallest scales. The most investigated solution is the cosmological model $\Lambda$ cold dark matter ($\Lambda$CDM), where GR is valid and two dark components are introduced, dark energy (DE) and dark matter (DM), to explain the $\sim$70\% and $\sim$25\% of the mass--energy budget of the Universe, respectively. An alternative approach is provided by modified gravity theories, where a departure of the gravity law from $\Lambda$CDM is assumed, and no dark components are included. This work presents refracted gravity (RG), a modified theory of gravity formulated in a classical way where the presence of DM is mimicked by a gravitational permittivity $\epsilon(\rho)$ monotonically increasing with the local mass density $\rho$, which causes the field lines to be refracted in small density environments. Specifically, the flatter the system the stronger the refraction effect and thus, the larger the mass discrepancy if interpreted in Newtonian gravity. RG presented several encouraging results in modelling the dynamics of disk and elliptical galaxies and the temperature profiles of the hot X-ray emitting gas in galaxy clusters and a covariant extension of the theory seems to be promising.}

\keyword{modified gravity; dark matter; galaxy dynamics; acceleration scale; scaling relations; galaxy clusters; cosmology} 

\begin{document}

\section{Introduction}
\label{sec:Introduction}

In the Universe, several pieces of evidence cannot be reconciled with general relativity (GR), if we only assume the presence of visible (baryonic) matter. A mass discrepancy of $\sim$80--90\% is observed from the largest scales (e.g., the cosmic microwave background (CMB)~\cite{Planck_2020}, the large-scale structure~\cite{Davis_1985,Springel_2006}, gravitational lensing effects where the Bullet Cluster example is worth mentioning~\cite{Markevitch_2006,Clowe_2006,Paraficz_2016},  light elements' abundances~\cite{Cyburt_2016}, and galaxy clusters dynamics~\cite{Zwicky_1933}) to the smallest scales (e.g., the flat trend of disk galaxies' rotation curves~\cite{Rubin_and_Ford_1970,Bosma_1978,Sanders_1990}). 

The most investigated solution to explain this phenomenology is to assume the presence of a nonbaryonic and cold, i.e., nonrelativistic at the epoch of decoupling from radiation, form of matter which only gravitationally interacts with baryonic matter: the cold dark matter (CDM)~\cite{Dodelson_1996}. However, the only presence of dark + baryonic matter is again not sufficient to account for all the pieces of evidence in the Universe, which suggest a further discrepancy of $\sim$70\%. The most studied cosmological model is the $\Lambda$CDM, which assumes GR as the gravity theory and introduces two dark constituents besides visible matter: dark energy (DE) and dark matter (DM), which can explain the $\sim$70\% and $\sim$25\% of the mass--energy budget of the Universe. Specifically, DE is an exotic fluid with negative pressure which justifies the accelerated expansion of the Universe, as observed from the Hubble diagram of Ia Supernovae (SNe Ia)~\cite{Kirshner_1996,Peebles_and_Ratra_2003}, and it can be identified with the cosmological constant $\Lambda$.

Even if $\Lambda$CDM can account for the majority of the observations in the Universe, it presents some problems, both on large and small scales. On large scales, we observe the cosmological constant problem~\cite{Weinberg_1989,Luo_2014}, the coincidence problem~\cite{Velten_2014}, and the tensions between the values of some cosmological parameters measured from probes of the late and early Universe~\cite{Planck_2020,Fleury_2013,Douspis_2019}. On small scales, we observe several discrepancies between $\Lambda$CDM simulations and observations. These generate the cusp/core, missing satellites, too-big-to-fail, and planes of satellite galaxies' problems (e.g., see~\cite{Del_Popolo_And_Le_Delliou_2017,deMartino_2020} for a review). On a galaxy scale, another issue related to the dynamics of galaxies is present in $\Lambda$CDM. This is the ``disk-halo conspiracy'', since the relative contributions of the baryonic stellar disk and the DM halo to the overall rotation curves of disk galaxies are degenerate to each other, that is, their contribution is not univocal. Several pieces of evidence suggest the ``maximum-disk hypothesis'' to solve this issue~\cite{van_Albada_and_Sancisi_1986,Sackett_1997,Courteau_and_Rix_1998,Bissantz_and_Gerhard_2002,Sellwood_and_Debattista_2014,McGaugh_and_Schombert_2015}, which consists in assuming that the baryonic stellar disk maximally contributes to the innermost part of the rotation curve. Yet, this assumption is only valid for high-surface-brightness (HSB) galaxies, whose rotation curves steeply rise in their inner regions and are modelled with a cuspy DM density profile~\cite{Dubinski_and_Carlberg_1991,Navarro_Frenk_and_White_1996}. On the contrary, low-surface-brightness (LSB) galaxies, generally dwarf and dwarf spheroidal (dSph) galaxies, are ones of the darkest objects at a galaxy scale in the Universe and they are DM-dominated as a result, even in their central regions: they have an inner velocity dispersion $\sigma \sim 10$~km~s$^{-1}$, an order of magnitude larger than the velocity dispersion $\sigma \sim 1$~km~s$^{-1}$ for systems with the same luminosity and scale radius ($\sim$100~pc) at equilibrium \cite{Mateo_Dwarfs_1998}. Their rotation curves, slowly rising toward their flat region, are described by a cored DM density profile and not by a maximum-disk model (e.g.,~\cite{Torrealba_2019}), which does not naturally emerge in $\Lambda$CDM cosmological simulations and gives rise to the cusp/core problem (e.g., ~\cite{Del_Popolo_And_Le_Delliou_2017,deMartino_2020}). Instead, there are systems with baryonic masses similar to those of dwarf galaxies, the globular clusters (GCs), which are nearly DM-free~\cite{Baumgardt_GCs_2005,Baumgardt_GCs_2009,Jordi_GCs_2009,Sollima_and_Nipoti_GCs_2010,Ibata_GCs_2011a,Ibata_GCs_2011b,Frank_GCs_2012}.

It is remarkable to cite also the presence of some observed regularities on a galaxy scale, which are hard to explain in a $\Lambda$CDM paradigm, where a stochastic merging process of structures' formation is invoked. Some of these regularities are very tight scaling relations between a property of the DM and of the baryonic matter in galaxies, which might be counter-intuitive since DM represents $\sim$90\% of the galaxy content, whereas baryonic matter only represents $\sim$10\%. Among these relations, we can mention (1) the baryonic Tully--Fisher relation (BTFR)~\cite{McGaugh_BTFR_2000}, (2) the mass discrepancy--acceleration relation (MDAR)~\cite{McGaugh_MDAR_2004}, and (3) the radial acceleration relation (RAR)~\cite{McGaugh_RAR_2016}. The three relations see the emergence of the same acceleration scale $a_0 \simeq 1.2 \times 10^{-10}$~m~s$^{-2}$, which, if expressed in natural units, is $a_0 \sim H_0 \sim \sqrt{\Lambda}$, suggesting a connection between the DM and DE sectors. These coincidences are even less intuitive in a $\Lambda$CDM context.

An alternative way to explain these discrepancies or regularities is provided by alternative theories of gravity, without the addition of any dark constituent. One of the most investigated modified theories of gravity formulated in a nonrelativistic way is modified Newtonian dynamics (MOND)~\cite{Milgrom_1983a,Milgrom_1983b,Milgrom_1983c}. MOND assumes a modification of the law of gravity dependent on the value of the background acceleration: when its value goes below the acceleration scale $a_0$, the gravitational field departs from Newtonian gravity, being subjected to a boost that mimics the effect of DM. MOND not only explains more intuitively but actually predicts several aspects of the dynamics of galaxies, such as the flatness of the rotation curves of disk galaxies and the three mentioned scaling relations.

Matsakos and Diaferio~\cite{Matsakos_and_Diaferio_RG_2016} proposed in 2016 a different approach. They formulated refracted gravity (RG), a classical modified theory of gravity which does not assume the presence of DM and is regulated by the value of the local mass density $\rho$, rather than of the acceleration $a$. The Poisson equation of RG is modified at first member by the presence of the gravitational permittivity, $\epsilon(\rho)$, a monotonic increasing function of $\rho$ which boosts the gravitational field in regions where the density goes below a critical value, reproducing the effect of DM in Newtonian gravity. A covariant version of RG was recently formulated~\cite{Sanna_CRG_2023} and it seems to describe both the DE and DM sectors with a single scalar field and to reproduce the Hubble diagram of SNe Ia.

RG has obtained some encouraging results in modelling the dynamics of galaxies and galaxy clusters, as well as on the covariant side. The paper develops as follows. Section~\ref{sec:RG} describes RG theory. Sections~\ref{sec:Disk_Galaxies}--\ref{sec:Covariant_RG} recap the main analyses and results obtained with RG in the field of disk galaxies, elliptical galaxies, galaxy clusters, and covariant RG, respectively. Section~\ref{sec:Discussion_And_Conclusions} discusses the future projects of RG and concludes the paper.

\section{Refracted Gravity}
\label{sec:RG}

RG is a theory of modified gravity formulated in a classical way, which can be interpreted in analogy to electrodynamics in matter~\cite{Matsakos_and_Diaferio_RG_2016}: as an electric field line suffers a change both in direction and in magnitude when it crosses a dielectric medium with a nonuniform permittivity, a gravitational field line suffers the same changes when it passes from a high-density to a low-density environment. This behaviour of the gravitational field is encoded in this modified Poisson equation (Equation (2.3) in~\cite{Matsakos_and_Diaferio_RG_2016}):
\begin{equation}
\label{eq:RG_Poisson_Equation}
\nabla \cdot \left[\epsilon(\rho)\nabla\phi\right] = 4\pi{G}\rho,
\end{equation}
where $\phi$ is the gravitational potential, and $\epsilon(\rho)$ is the gravitational permittivity. The following asymptotic limits are adopted for the gravitational permittivity (Equation (2.5) in~\cite{Matsakos_and_Diaferio_RG_2016}):
\begin{equation}
    \label{eq:epsilonRGasymptreg}
    	\epsilon(\rho) =
    	\begin{cases}
    		1, & \rho \gg \rho_{\rm c}\\
    		\epsilon_0, & \rho \ll \rho_{\rm c},
    	\end{cases}
\end{equation}
where $0 < \epsilon_0 \leq 1$, and $\rho_{\rm c}$ are the gravitational permittivity in vacuum and the critical density, respectively, two of the three free parameters of the theory. When $\rho \gg \rho_{\rm c}$, the Newtonian Poisson equation is recovered (Equation (2.7) in~\cite{Matsakos_and_Diaferio_RG_2016}):
\begin{equation}
\label{eq:Newtonian_Poisson_Equation}
\nabla^2\phi_{\rm N} = 4\pi{G}\rho,
\end{equation}
whereas, when $\rho \ll \rho_{\rm c}$, we are in a fully RG regime.

\begin{figure}[H]
 \hspace{-6pt}    		\includegraphics[scale=0.50]%
   	{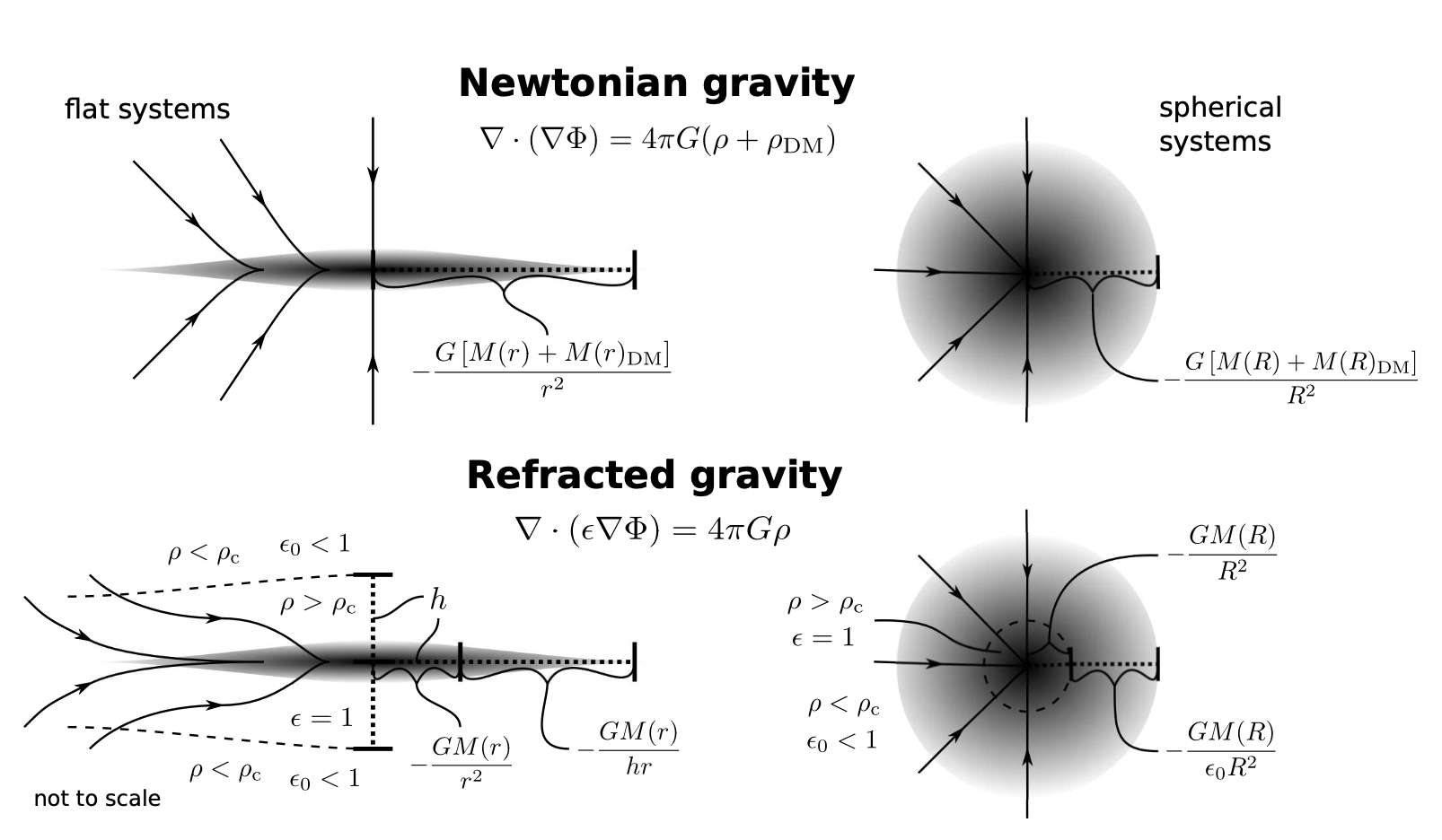}
    	\caption{Comparison between the behaviour of Newtonian ({\bf{top panels}}) and RG ({\bf bottom panels}) gravitational field lines for flat ({\bf left panels}) and spherical ({\bf right panels}) systems. The formulae reported for flattened systems, in the left part, are equivalent formulae for spherical systems to approximately illustrate the dependence of the
  field on the distance from the galaxy centre. The figure is reproduced from Figure 16 in~\cite{Matsakos_and_Diaferio_RG_2016}.}	\label{fig:Newtonian_RG_Field_Flat_Spherical_Systems}
\end{figure}

The RG gravitational field has a different behaviour for spherical and flat systems. For spherical systems, the RG Poisson equation (Equation~\eqref{eq:RG_Poisson_Equation}) reduces to (Equation (3) in~\cite{Cesare_2022}):
\begin{equation}
    \label{eq:RGfieldsph}
    \frac{\partial\phi}{\partial r} = \frac{1}{\epsilon(\rho)} \frac{G M(< r)}{r^2} = \frac{1}{\epsilon(\rho)} \frac{\partial\phi_{\rm N}}{\partial r},
\end{equation}
where $M(< r)$ is the mass of the spherical system enclosed within the spherical radius $r$, and $\partial\phi_{\rm N}/\partial{r}$ is Newtonian gravitational field. In this case, the RG field direction and $r$-dependence remain Newtonian, and the RG field magnitude increases compared to the Newtonian one where $\rho \ll \rho_{\rm c}$, i.e., where $\epsilon(\rho) \rightarrow \epsilon_0$ (right panels of Figure~\ref{fig:Newtonian_RG_Field_Flat_Spherical_Systems}). If we consider a solid sphere immersed in vacuum ($\rho = 0$) with radius $r_{\rm s}$ and uniform density $\rho_{\rm s} \gg \rho_{\rm c}$, the RG gravitational field inside and outside the sphere is given by~\cite{Matsakos_and_Diaferio_RG_2016}:
\begin{equation}
    \label{eq:g_RG_In_Out_Solid_Sphere_In_Vacuum}
    	g =
    	\begin{cases}
    		g_{\rm N}, & r < r_{\rm s}\\
    		\frac{g_{\rm N}}{\epsilon_0}, & r > r_{\rm s},
    	\end{cases}
\end{equation}
\textcolor{black}{where $g = \partial\phi/\partial{R}$ and $g_{\rm N} = \partial\phi_{\rm N}/\partial{R}$ are the RG and Newtonian gravitational fields, respectively. The gravitational field has the Newtonian direction everywhere but is enhanced by $1/\epsilon_0$ over the Newtonian field outside the sphere.}

The analogy with electrodynamics in matter, i.e., the refraction of the field lines where the density decreases, is observed for flattened systems. Expanding the left-hand side of Equation~\eqref{eq:RG_Poisson_Equation}, we obtain (Equation (4) in~\cite{Cesare_2020b}):
\begin{equation}
    \label{eq:PoissonRGexpanded1stLHS}
    \frac{\partial \epsilon}{\partial \rho} \nabla\rho \cdot \nabla\phi + \epsilon(\rho)\nabla^2\phi = 4\pi G \rho.
\end{equation}

{In} 
 this configuration, the RG field depends both on the density $\rho$ (second term on the left-hand side of Equation~\eqref{eq:PoissonRGexpanded1stLHS}) as for spherical systems, and on its gradient (first term on the left-hand side of Equation~\eqref{eq:PoissonRGexpanded1stLHS}). In particular, the term ``$\frac{\partial \epsilon}{\partial \rho} \nabla\rho \cdot \nabla\phi$'' is different from zero in nonspherical configurations and causes the field lines to refract. The acceleration boost in the external regions of disk galaxies that in Newtonian gravity is explained by the presence of DM, is explained in RG by the refraction of the field lines toward the equatorial plane of the disk caused by the low-density regions above and below the disk plane (left panels of Figure~\ref{fig:Newtonian_RG_Field_Flat_Spherical_Systems}). Following Equation~\eqref{eq:PoissonRGexpanded1stLHS}, RG predicts that the flatter the system, the larger the mass discrepancy if interpreted in Newtonian gravity. This might explain a positive correlation between ellipticity and DM content of elliptical galaxies~\cite{Deur_2014,Deur_2020} and the different DM quantity of GCs and dwarf galaxies, where the former are nearly spherical and DM-free (e.g.,~\cite{Sollima_and_Nipoti_GCs_2010}) and the latter are flatter and one of the darkest objects in the Universe (e.g.,~\cite{Mateo_Dwarfs_1998}) (see Section~\ref{sec:Introduction}). It has to be specified that the formulae reported for flattened systems in the left panels of Figure~\ref{fig:Newtonian_RG_Field_Flat_Spherical_Systems} are equivalent formulae for spherical systems to approximately illustrate the dependence of the field on the distance from the galaxy centre.
 \vspace{-12pt}

Where $\rho \ll \rho_{\rm c}$ in flat systems, e.g., in the outskirts of disk galaxies, the radial gravitational field assumes the asymptotic trend (Equation (B.23) in~\cite{Cesare_2020b}): 
\textcolor{black}{
\begin{equation}
    \label{eq:RG_Field_Radial_Asymptotic_Limit}
    \frac{\partial\phi}{\partial{R}} = \sqrt{\Big\vert\frac{\partial\phi_{\rm N}}{\partial{R}}\Big\vert a_0} \propto R^{-1}
\end{equation}
}
\hspace{-3pt}where $a_0$ is the MOND critical acceleration set by the normalisation of the BTFR \cite{McGaugh_2012}:
\begin{equation}
    \label{eq:BTFR}
    v_{\rm f}^4 = Ga_0M,
\end{equation}
where $v_{\rm f}$ is the asymptotic value of the flat part of the rotation curve of disk galaxies and $M$ is the total baryonic mass of the disk galaxy. \textcolor{black}{ Specifically, Matsakos and Diaferio~\cite{Matsakos_and_Diaferio_RG_2016} derived the relation $v_{\rm f}^4 = GbM$ ({\textcolor{black}{Matsakos and Diaferio~\cite{Matsakos_and_Diaferio_RG_2016} used the lowercase font for the baryonic mass $m$, whereas I adopt the uppercase font for the same symbol across the paper}}), where they put the acceleration parameter $b$ in relation with the MOND acceleration scale $a_0$ to be consistent with the normalisation of the observed BTFR~\cite{McGaugh_2012,Famaey_and_McGaugh_2012}.} In this regime, the RG field trend deviates from the Newtonian inverse square law. This asymptotic limit coincides with the MOND asymptotic limit for the gravitational field where $a \ll a_0$, which might indicate that RG shares the majority of MOND's successes on a galaxy scale. The difference between RG and MOND is better observed in spherical systems, where the field trend in RG is Newtonian ($\propto R^{-2}$) even where $\rho \ll \rho_{\rm c}$, preserving the Gauss theorem, but the field trend in MOND is $\propto R^{-1}$ where $a \ll a_0$, as in flattened systems.

\textcolor{black}{In the regions of flat systems where $\rho \ll \rho_{\rm c}$, besides the asymptotic limit given by Equation~\eqref{eq:RG_Field_Radial_Asymptotic_Limit} for the radial gravitational field, the following asymptotic limit for the vertical gravitational field holds (Equation (B.25) in~\cite{Cesare_2020b}):}
\textcolor{black}{
\begin{equation}
    \label{eq:RG_Field_Vertical_Asymptotic_Limit}
    \frac{\partial\phi}{\partial{z}} = 0,
\end{equation}
}
\hspace{-3pt}{which means that the field is redirected parallel to the equatorial plane of the flat system at the height where condition~\eqref{eq:RG_Field_Vertical_Asymptotic_Limit} is reached. If we consider a solid disk immersed in vacuum ($\rho = 0$) with radius $R_{\rm d}$, height $h_{\rm d}$ from its equatorial plane (2$h_{\rm d}$ of total height), and uniform density $\rho_{\rm d} \gg \rho_{\rm c}$, its RG gravitational field remains Newtonian within the disk and reaches the asymptotic limits~\eqref{eq:RG_Field_Radial_Asymptotic_Limit} and~\eqref{eq:RG_Field_Vertical_Asymptotic_Limit} outside the disk. Specifically, condition~\eqref{eq:RG_Field_Vertical_Asymptotic_Limit} is accomplished at distance $z = \pm h_{\rm d}$ from the equatorial plane of the disk.}

\begin{figure}[H]

    		\includegraphics[scale=1.40]%
    		{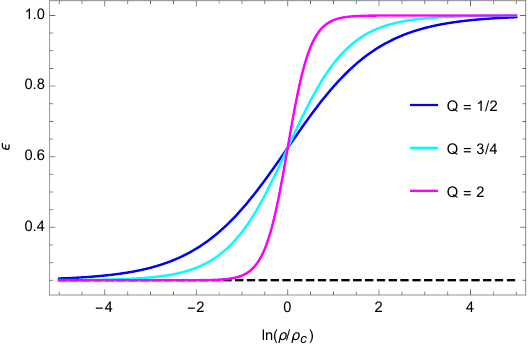}
    		\caption{Gravitational permittivity $\epsilon(\rho)$ adopted in RG (Equation~\eqref{eq:epsilon}) for three different values of the RG free parameter $Q$ and for $\epsilon_0 = 0.25$. The larger the value of $Q$, the steeper the transition between the two asymptotic limits of Equation~\eqref{eq:epsilonRGasymptreg}. The black dashed line represents $\epsilon_0 = 0.25$. The figure is reproduced from Figure 1 in~\cite{Cesare_2020b}. Credit: Cesare V., Diaferio A., Matsakos T., and Angus G., A\&A, 637, A70, 2020, reproduced with permission
$\copyright$ ESO.}
    		\label{fig:epsilon}
	
\end{figure}

Different functions can be adopted for the gravitational permittivity, such that it accomplishes the asymptotic limits of Equation~\eqref{eq:epsilonRGasymptreg}. For all the analyses performed with RG, the following smooth step function of the mass density $\rho$ was employed \textcolor{black}{(Equation (4.1) in~\cite{Matsakos_and_Diaferio_RG_2016}), Equation (5) in~\cite{Cesare_2020b}, and Equation (4) in~\cite{Cesare_2022}}:
\begin{equation}
    \label{eq:epsilon}
    \epsilon(\rho)=\epsilon_0+(1-\epsilon_0)\frac{1}{2}\left\{\tanh\left[\text{ln}\left(\frac{\rho}{\rho_\mathrm{c}}\right)^Q\right]+1\right\},
\end{equation}
represented in Figure~\ref{fig:epsilon}. Besides $\epsilon_0$ and $\rho_{\rm c}$, a third free parameter, $Q$, is present in RG, where $Q$ regulates the steepness of the transition between the two asymptotic limits of Equation~\eqref{eq:epsilonRGasymptreg}. In Figure~\ref{fig:epsilon}, Equation~\eqref{eq:epsilon} is represented for three values of $Q$, showing that the larger its value, the steeper the transition between the two asymptotic regimes of Equation~\eqref{eq:epsilonRGasymptreg}. The three RG free parameters are supposed to be universal.

\section{Dynamics of Disk Galaxies}
\label{sec:Disk_Galaxies}

Matsakos and Diaferio~\cite{Matsakos_and_Diaferio_RG_2016} presented some preliminary encouraging results, according to which RG properly models the BTFR (Equation~\eqref{eq:BTFR}) \textcolor{black}{and the MDAR} of galaxies, and the rotation curves of the HSB disk galaxy NGC 6946 and LSB disk galaxy NGC 1560. \mbox{Cesare et~al.~\cite{Cesare_2020b}} demonstrated that RG can properly model the dynamics of disk galaxies. For their analysis, they considered the rotation curves and the vertical velocity dispersion's radial profiles of 30 disk galaxies in the DiskMass Survey (DMS)~\cite{Bershady_DMS_I_2010a} at low redshift. They chose this sample since these galaxies are close to being face-on, which allowed the measurement of both their rotation curves and vertical velocity dispersions. In this way, Cesare et al.~\cite{Cesare_2020b} could obtain a single constraint of the RG parameters from two kinematic profiles taken at the same time rather than from the rotation curves alone, which provided a more robust analysis. These results are also summarised in~\cite{Cesare_2021}.

Cesare et al.~\cite{Cesare_2020b} modelled with RG (1) the rotation curves alone of each DMS galaxy, (2) the rotation curves and vertical velocity dispersions at the same time of each DMS galaxy, (3) the rotation curves and vertical velocity dispersions of all 30 DMS galaxies at the same time, and (4) the RAR of DMS galaxies.

\subsection{Mass Model}
\label{sec:Disk_Galaxies_Mass_Model}
Cesare et al.~\cite{Cesare_2020b} modelled the baryonic mass density profile $\rho$, which generates the total gravitational potential $\phi$ \textcolor{black}{in the hypothesis that DM is not present}, with (1) an axisymmetric exponential disk for the stars, (2) a spherical stellar bulge, and (3) two razor-thin disks for the atomic and molecular gas. The 3D mass density of the stellar disk was modelled with an exponential profile of this kind \textcolor{black}{(Equation (7) in~\cite{Cesare_2021})}:
\begin{equation}
    \label{eq:rho_stellar_disk}
    \rho_{\rm d}(R,z) = \frac{\Upsilon{I_{\rm d}(R)}}{2h_z}\exp\left(-\frac{|z|}{h_z}\right),
\end{equation}
where $I_{\rm d}(R)$ is the surface brightness radial profile of the stellar disk, $R$ and $z$ are the cylindrical coordinates ($R$ is the radius projected on the sky, oriented along the major axis of the disk, and $z$ is the vertical coordinate, oriented perpendicular to the disk equatorial plane), and $\Upsilon$ and $h_z$ are the stellar mass-to-light ratio and the disk scale height, two of the five free parameters of the dynamical model adopted by~\cite{Cesare_2020b}. 

The surface brightness of the disk was modelled with a linear interpolation of the measured surface brightness data to catch the specific features of the luminous matter distribution, which generally correspond to features in the rotation curve, following ``Renzo's rule''~\cite{Sancisi_2004}. The disk surface brightness in the innermost galaxy regions, where the bulge is dominant, was modelled with an exponential profile~\cite{de_Vaucouleurs_1959,Freeman_1970} \textcolor{black}{(Equation (A.1) in~\cite{Cesare_2020b})}:
\begin{equation}
\label{eq:I_disk_inner_galaxy_interpolation}
I_{\rm d}(R) = I_{{\rm d},0}\exp\left(-\frac{R}{h_R}\right),
\end{equation}
obtained by fitting the surface brightness points of the outermost stellar disk, where the stellar disk alone dominates.

DMS galaxies are disk-dominated, the average stellar bulge-over-total luminosity ratio in the $K$-band being around $0.09$. For this reason, the bulge was approximated as a sphere, introducing negligible systematic errors. The 3D mass density of the bulge was modelled with this Abel integral \textcolor{black}{(Equation (9) in~\cite{Cesare_2021})}:
\begin{equation}
    \label{eq:rho_bulge}
    \rho_{\rm b}(r) = -\frac{\Upsilon}{\pi}\int^{+\infty}_r\frac{{\rm d}I_{\rm b}(R)}{{\rm d}R}\frac{{\rm d}R}{\sqrt{R^2-r^2}},
\end{equation}
where $I_{\rm b}(R)$ is the surface brightness radial profile of the bulge, and 
\begin{equation}
    \label{eq:3D_radius_r}
    r = \sqrt{R^2 + z^2}
\end{equation}
is the 3D radius. The bulge contribution being subdominant, Cesare et al.~\cite{Cesare_2020b} assumed the mass-to-light ratio of the bulge equal to the mass-to-light ratio of the disk, $\Upsilon$, without introducing an additional degree of freedom. They modelled the surface brightness of the bulge with a Sérsic spherical profile~\cite{Sersic_1963} \textcolor{black}{(Equation (A.2) in~\cite{Cesare_2020b})}:
\begin{equation}
\label{eq:I_bulge_real}
I_{\rm b}(R) = I_{\rm e}\exp\left\{-7.67\left[\left(\frac{R}{R_{\rm e}}\right)^{1/n_{\rm s}} - 1\right]\right\}.
\end{equation}

{To} 
 account for the seeing, which is not negligible since the measurements were taken from the ground with the 3.5 m diameter Calar Alto Observatory, they modelled the observed surface brightness of the bulge with Equation~\eqref{eq:I_bulge_real} convolved with a Gaussian point spread~function.

The 3D mass density of the atomic and molecular gas were modelled with two razor-thin disks \textcolor{black}{(Equation (A.14) in~\cite{Cesare_2020b})}:
\begin{equation}
    \label{eq:rho_atomic_gas}
    \rho_{\rm atom}(R,z) = \Sigma_{\rm atom}(R)\delta(z),
\end{equation}
and
\begin{equation}
    \label{eq:rho_molecular_gas}
    \rho_{\rm mol}(R,z) = \Sigma_{\rm mol}(R)\delta(z),
\end{equation}
where $\Sigma_{\rm atom}(R)$ and $\Sigma_{\rm mol}(R)$ are the surface mass densities of the atomic and molecular gas, respectively, which Cesare et al.~\cite{Cesare_2020b} modelled with a linear interpolation of the data \textcolor{black}{as for the stellar disk, to catch the features of the rotation curve}, and $\delta(z)$ is the Dirac $\delta$ function.

\textls[-20]{The total 3D mass density profile is given by summing Equations~\eqref{eq:rho_stellar_disk},~\eqref{eq:rho_bulge},~\eqref{eq:rho_atomic_gas} and~\eqref{eq:rho_molecular_gas}} \textcolor{black}{(Equation (11) in~\cite{Cesare_2021})}:
\begin{equation}
    \label{eq:rho_total}
    \rho(R,z) = \rho_{\rm d}(R,z) + \rho_{\rm b}(r) + \rho_{\rm atom}(R,z) + \rho_{\rm mol}(R,z).
\end{equation}

\subsection{Dynamical Model}
\label{sec:Disk_Galaxies_Dynamical_Model}

From the total 3D mass density profile given by Equation~\eqref{eq:rho_total}, Cesare et al.~\cite{Cesare_2020b} derived the RG gravitational potential by solving the RG Poisson Equation~\eqref{eq:RG_Poisson_Equation}. To solve the RG Poisson equation, they used an iterative Poisson solver based on the successive-over-relaxation method, given by combining the Jacobi and Gauss--Seidel methods~\cite{Young_1954}. From the RG potential, they derived the models for the rotation curve, $v(R)$ \textcolor{black}{(Equation (6) in~\cite{Cesare_2020b})},
\begin{equation}
    \label{eq:Rotation_Curve}
    v(R,z=0) = \sqrt{R\frac{\partial\phi(R,z)}{\partial{R}}},
\end{equation}
\textcolor{black}{and the vertical velocity dispersion radial profile, $\sigma_z(R)$ (Equation (13) in~\cite{Cesare_2020b}),}
\begin{equation}
\label{eq:Vertical_Velocity_Dispersion_Radial_Profile}
\sigma_z(R)=\sqrt{\frac{1}{h_z}\int_0^{+\infty}\left[\int_z^{+\infty}\exp\left(-\frac{|z'|}{h_z}\right)\frac{\partial\phi(R,z')}{\partial z'}dz'\right]dz},
\end{equation}
where Equation~\eqref{eq:Vertical_Velocity_Dispersion_Radial_Profile} is the vertical velocity dispersion for a vertically decreasing exponential disk whose density follows Equation~\eqref{eq:rho_stellar_disk}.

The dynamical model has five free parameters: the stellar mass-to-light ratio $\Upsilon$, the stellar disk scale height $h_z$, and the three RG parameters, $\epsilon_0$, $Q$, and $\rho_{\rm c}$. They estimated the free parameters of the model firstly from the rotation curve alone of each galaxy and then from the rotation curve and the vertical velocity dispersion taken at the same time of each galaxy with a Monte Carlo Markov Chain (MCMC) algorithm with a Metropolis--Hastings acceptance criterion. They adopted the following priors for the free parameters of the model. For $\Upsilon$, they adopted a Gaussian prior based on the stellar population synthesis (SPS) model of Bell and de Jong~\cite{Bell_And_deJong_2001}, keeping into account that the surface brightness was measured in the $K$-band, where the Gaussian tail for $\Upsilon < 0$ was set to zero. For $h_z$, they adopted a Gaussian prior peaking at $h_{z,{\rm SR}}$, a scale height obtained with the following relation, derived from a combined sample of 60 edge-on late-type galaxies by Bershady, et~al.~\cite{Bershady_DMS_II_2010b} \textcolor{black}{(Equation~(1) in~\cite{Bershady_DMS_II_2010b} and Equation (A.12) in~\cite{Cesare_2020b})}:
\begin{equation}
\label{eq:hzhR}
\log_{10}\left(\frac{h_R}{h_{z,\text{SR}}}\right)=0.367\log_{10}\left(\frac{h_R}{\text{kpc}}\right)+0.708 \pm 0.095 \; ,
\end{equation}
where the term $\pm 0.095$ indicates the $\sim$25\% intrinsic scatter. Estimating $h_{z,{\rm SR}}$ was essential to have a comparison reference for DMS galaxies where, due to their nearly face-on configuration, their $h_z$ could not be directly measured. For the three RG parameters, they adopted a flat prior in the following intervals: $[0.10,1.00]$ for $\epsilon_0$, $[0.01,2.00]$ for $Q$, and $[-27.00,-23.00]$ for $\log_{10}[\rho_{\rm c} ({\rm g}/{\rm cm}^3)]$.

\subsection{Results}
\label{sec:Disk_Galaxies_Results}

\begin{figure}[H]
 \hspace{-6pt}   		\includegraphics[scale=0.75]%
    		{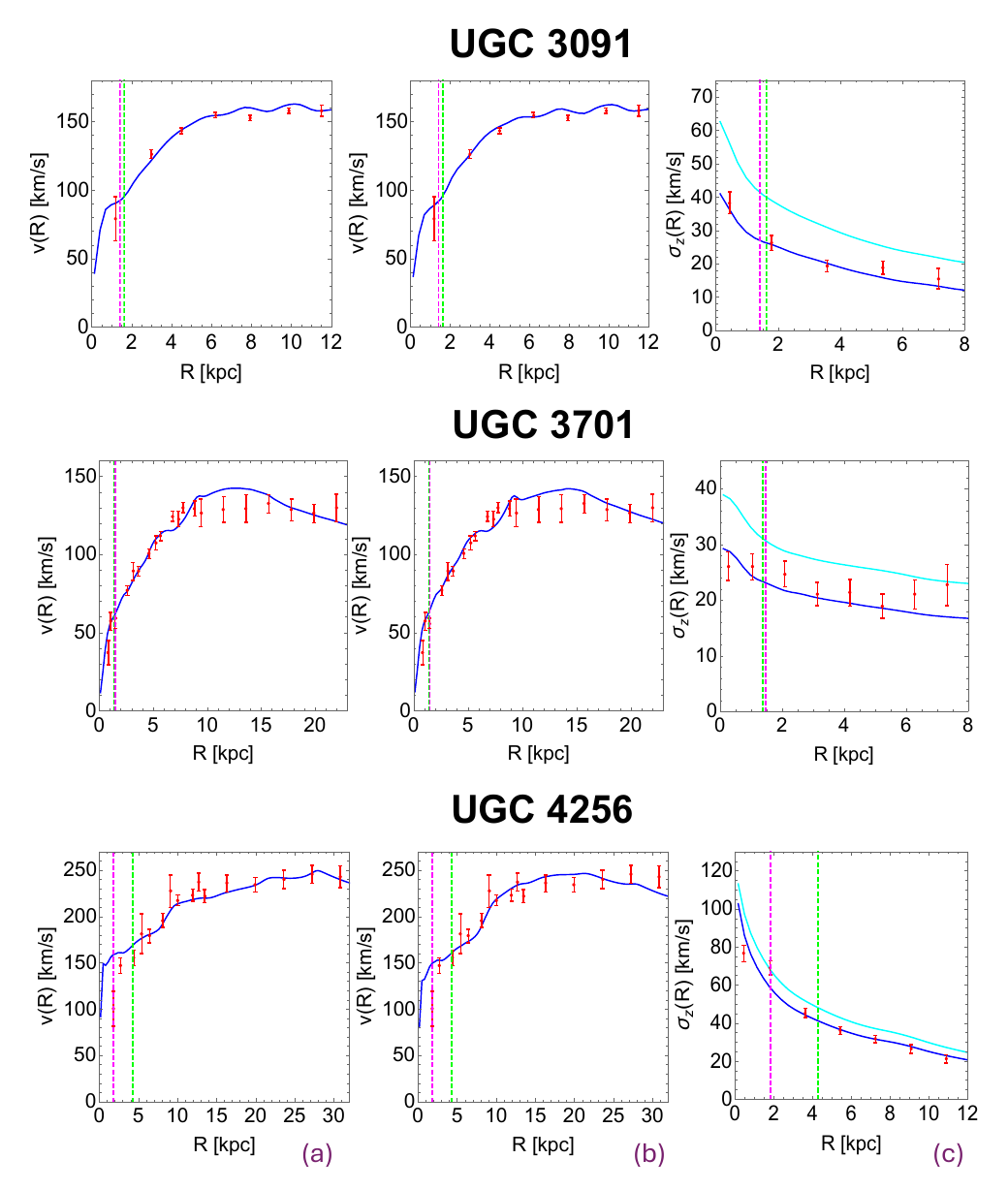}
    		\caption{RG models (blue solid lines) of the rotation curves and the vertical velocity dispersion profiles against the DMS data (red dots with error bars) for three DMS galaxies (UGC 3091, UGC 3701, and UGC 4256). Panels ({\bf a}): Rotation curves computed with the parameters found by modelling the rotation curves alone of each galaxy. Panels ({\bf b}): Rotation curves computed with the parameters found by modelling the rotation curves and vertical velocity dispersion profiles of each galaxy at the same time. Panels ({\bf c}): Vertical velocity dispersion profiles computed with the parameters found by modelling the rotation curves and vertical velocity dispersion profiles of each galaxy at the same time. The cyan solid lines are the RG vertical velocity dispersion profiles calculated with the same parameters as the blue curves but with disk-scale heights $h_z = h_{z,{\rm SR}}$ (Equation~\eqref{eq:hzhR}), and the dashed magenta and green vertical lines show the bulge effective radius ($R_{\rm e}$ in Equation~\eqref{eq:I_bulge_real}) and the bulge radius adopted in the disk-bulge decomposition for the surface brightness modelling, respectively. The figure is readapted from Figures D.1 and D.2 in~\cite{Cesare_2020b}. Credit: Cesare V., Diaferio A., Matsakos T., and Angus G., A\&A, 637, A70, 2020, reproduced with permission
$\copyright$ ESO.}
    		\label{fig:Results_RG_Disk_Galaxies}
\end{figure}

RG modelled the kinematic profiles of DMS galaxies with sensible parameters. Figure~\ref{fig:Results_RG_Disk_Galaxies} shows the RG models (blue solid lines) of the rotation curves (panels (a) and (b)) and the vertical velocity dispersion profiles (panels (c)) against the DMS data (red dots with error bars) for three DMS galaxies (UGC 3091, UGC 3701, and UGC 4256). The blue curves for the rotation curves in panels (a) were computed with the parameters estimated from the rotation curves alone of each galaxy. The blue curves for the rotation curves in panels (b) were computed with the parameters found by modelling the rotation curves and vertical velocity dispersion profiles of each galaxy at the same time. \textcolor{black}{With the same parameters, the vertical velocity dispersion profiles were computed in panels (c).} In panels (c), the cyan solid lines are the RG vertical velocity dispersions calculated with the same parameters as the blue curves but with disk-scale heights $h_z$ set to $h_{z,{\rm SR}}$ (Equation~\eqref{eq:hzhR}). In all panels, the dashed magenta vertical lines identify the bulge effective radius ($R_{\rm e}$ in Equation~\eqref{eq:I_bulge_real}), and the dashed green vertical lines represent the bulge radius adopted in the disk-bulge decomposition for the surface brightness modelling (see Section~\ref{sec:Disk_Galaxies_Mass_Model}). We can see that RG properly modelled both kinematic profiles and that the substructures of the rotation curves were properly reproduced following Renzo's rule (see Section~\ref{sec:Disk_Galaxies_Mass_Model}).

Both in the only-rotation curve and in the combined rotation curve + vertical velocity dispersion analyses, Cesare et al.~\cite{Cesare_2020b} found mass-to-light ratios $\Upsilon$ generally consistent with the SPS model of Bell and de Jong~\cite{Bell_And_deJong_2001}. In the only-rotation curve analysis, the estimated $h_z$'s were generally consistent with the $h_{z,{\rm SR}}$ derived from Equation~\eqref{eq:hzhR}. Instead, in the combined analysis, the estimated $h_z$'s were systematically smaller than the correspondent $h_{z,{\rm SR}}$. Angus et al.~\cite{Angus_2015} found a similar result by modelling the DMS galaxies with QUMOND~\cite{Milgrom_2010}, a modified gravity version of MOND theory. Yet, this result seems to not indicate a flaw in the two modified theories of gravity but instead, an observational bias. Indeed, the disk scale height $h_z$ found by Cesare et al.~\cite{Cesare_2020b} and by Angus et al.~\cite{Angus_2015} was estimated from the vertical velocity dispersions, which are measured with spectroscopy from a signal derived from a mixed stellar population of young and old stars, where the signal from the young population is dominant~\cite{Aniyan_2016}. The young stellar population distributes in a disk thinner and with a smaller vertical velocity dispersion $\sigma_z$ than the old stellar population~\cite{Aniyan_2016}. Instead, the disk scale heights of edge-on galaxies, from which Equation~\eqref{eq:hzhR} is derived to determine $h_{z,{\rm SR}}$, are directly measured from near-infrared photometry, whose signal is dominated by the old stellar population~\cite{Aniyan_2016}, having higher $h_z$ and $\sigma_z$ than the young stellar population. It is possible to simulate the vertical velocity dispersion of the old stellar population by artificially increasing the vertical velocity dispersion of the young stellar population by a proper factor. Cesare et al.~\cite{Cesare_2020b} computed this factor, which was equal to $1.63$, in agreement with the work of Aniyan et al.~\cite{Aniyan_2016}. Cesare et al.~\cite{Cesare_2020b} artificially increased this factor to determine the vertical velocity dispersions of five DMS galaxies, obtaining disk scale heights in agreement with the correspondent $h_{z,{\rm SR}}$. 

\begin{figure}[H]
    		\includegraphics[scale=0.80]%
    		{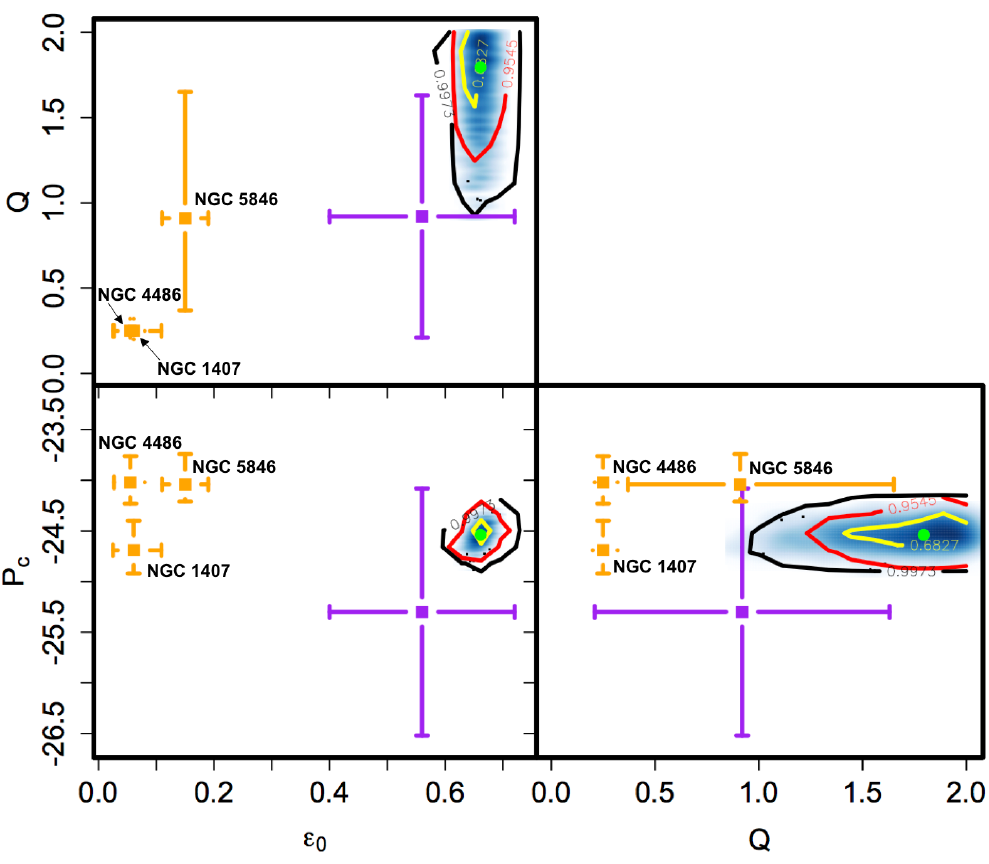}
    		\caption{Free parameters of the RG gravitational permittivity $\epsilon(\rho)$ estimated from the DMS and the three elliptical E0 galaxies. The purple squares with error bars represent the means of the permittivity parameters estimated at the same time from the rotation curves and the vertical velocity dispersion profiles of each DMS galaxy. The light blue shaded areas indicate the posterior distributions of the unique combination of RG parameters estimated from the entire DMS sample at the same time, where their medians and 1$\sigma$, 2$\sigma$, and 3$\sigma$ levels are shown as green dots and yellow, red, and black contours, respectively. The orange squares with error bars are the permittivity parameters derived from each elliptical E0 galaxy. The figure is reproduced from Figure 9 in~\cite{Cesare_2022}. Credit: Cesare V., Diaferio A., and Matsakos T., A\&A, 657, A133, 2022, reproduced with permission
$\copyright$ ESO.}
    \label{fig:RG_Parameters_Disk_Galaxies_Elliptical_Galaxies}
\end{figure}

The three RG parameters derived from the rotation curves and the vertical velocity dispersions of each galaxy are in agreement with each other: their average errors are larger than the standard deviation of their distributions. This suggests their universality, as should be from the theory formulation. To better test their universality, Cesare et al.~\cite{Cesare_2020b} estimated a unique combination of RG free parameters from the two kinematic profiles of all 30 galaxies taken at the same time, to verify their agreement with the average of the RG parameters found from the rotation curves and the vertical velocity dispersions of each galaxy. \textls[-15]{The mean RG parameters derived from the single galaxies with their mean errors were $\epsilon_{0,{\rm Mean,DMS}} = 0.56 \pm 0.16$, $Q_{\rm Mean,DMS} = 0.92 \pm 0.71$, $\log_{10}[\rho_{\rm c}\text{(g/cm}^3)]_{\rm Mean,DMS} = -25.30 \pm 1.22$ (purple squares with error bars in Figure~\ref{fig:RG_Parameters_Disk_Galaxies_Elliptical_Galaxies})}, whereas the unique combination of free parameters was $\epsilon_{0,{\rm Unique,DMS}} = 0.661^{+0.007}_{-0.007}$, $Q_{\rm Unique,DMS} = 1.79^{+0.14}_{-0.26}$, $\log_{10}[\rho_{\rm c}\text{(g/cm}^3)]_{\rm Unique,DMS} = -24.54^{+0.08}_{-0.07}$, where their estimates and error bars were the medians (green dots in Figure~\ref{fig:RG_Parameters_Disk_Galaxies_Elliptical_Galaxies}) and the $15.9$ and $84.1$ percentiles of their posterior distributions (blue shaded areas in Figure~\ref{fig:RG_Parameters_Disk_Galaxies_Elliptical_Galaxies}). In Figure~\ref{fig:RG_Parameters_Disk_Galaxies_Elliptical_Galaxies}, the yellow, red, and black contours show the 1$\sigma$, 2$\sigma$, and 3$\sigma$ levels. The code used to find this unique combination of RG parameters was written in C++ and parallelised with OpenMP. It is publicly available on GitHub under the name ``astroMP'' (\url{https://github.com/alpha-unito/astroMP}, accessed on 7 November 2019) and it is described in~\cite{Cesare_2020a,Aldinucci_2021}.
\vspace{-8pt}

The unique combination of RG parameters was found with an approximate procedure, in which the mass-to-light ratios and the disk scale heights were kept fixed to the values found from the single galaxies. The fact that this unique combination of free parameters provided a limited worsening of the agreement between the RG models and the kinematic data and that the unique $\epsilon_0$, $Q$, and $\log_{10}[\rho_{\rm c}\text{(g/cm}^3)]$ agreed with the correspondent RG parameters estimated from the single galaxies within $0.63$$\sigma$, $1.19$$\sigma$, and $0.62$$\sigma$, respectively, further enforced the hypothesis of the universality of these parameters. With a proper exploration of the 63-dimensional parameter space for this galaxy sample (one $\Upsilon$ and one $h_z$ for each of the 30 galaxies and three RG parameters for all the galaxies), a better consistency between model and measurements and a more precise estimate of the unique combination of RG parameters might be achieved.

With the parameters obtained from the two kinematic profiles of each galaxy, Cesare et al.~\cite{Cesare_2020b} also modelled the RAR of the DMS galaxies. McGaugh et al.~\cite{McGaugh_RAR_2016} fitted the RAR of 153 edge-on disk galaxies in the Spitzer Photometry and Accurate Rotation Curves (SPARC) sample~\cite{Lelli_SPARC_2016b} with this relation \textcolor{black}{(Equation (4) in~\cite{McGaugh_RAR_2016} and Equation (20) in~\cite{Cesare_2020b})}:
\begin{equation}
\label{eq:fittedRAR}
g_\text{obs}(R)=\frac{g_\text{bar}(R)}{1-\exp\left(-\sqrt{\frac{g_\text{bar}(R)}{g_\dagger}}\right)},
\end{equation}
where $g_{\rm obs}(R)$ is the observed centripetal acceleration derived from the measured rotation curve $v_{\rm obs}(R)$ \textcolor{black}{(Equation (19) in~\cite{Cesare_2020b})}, 
\begin{equation}
    \label{eq:gobs}
    g_{\rm obs}(R) = \frac{v^2_{\rm obs}(R)}{R},
\end{equation}
$g_{\rm bar}(R)$ is the Newtonian acceleration due to baryonic matter alone, and $g_\dagger$ is the only free parameter of the model. This relation properly interpolated the SPARC data with $g_\dagger = [1.20 \pm 0.02 {\rm\text{ }(random)} \pm 0.24 {\rm \text{ }(systematic)}] \times 10^{-10}$~m~s$^{-2}$, which agrees with the MOND acceleration scale $a_0$ within 1$\sigma$. 

\begin{figure}[H]
    		\includegraphics[scale=1.20]%
    		{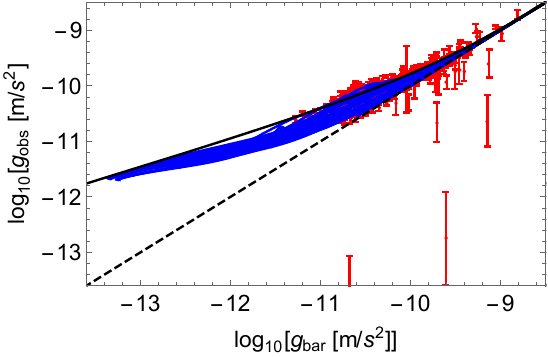}
    		\caption{RAR models calculated with RG (blue solid lines) with the parameters obtained from the MCMC analysis of the rotation curves and vertical velocity dispersion profiles of each galaxy at the same time, superimposed onto the RAR of DMS data (red points with error bars). The black solid line is the RAR fitting relation (Equation~\eqref{eq:fittedRAR}) obtained by McGaugh et al.~\cite{McGaugh_RAR_2016} from SPARC galaxies and the black dashed line is the $g_{\rm obs} = g_{\rm bar}$ relation, plotted as a reference. The figure is readapted from Figure 11 in~\cite{Cesare_2020b}. Credit: Cesare V., Diaferio A., Matsakos T., and Angus G., A\&A, 637, A70, 2020, reproduced with permission
$\copyright$ ESO.}
    		\label{fig:RAR_RG}
\end{figure}

The RG models of the RAR for the DMS galaxies (blue solid lines in Figure~\ref{fig:RAR_RG}) properly interpolated the measured RAR of the DMS data (red dots with error bars in Figure~\ref{fig:RAR_RG}) and reproduced the asymptotic limits of Equation~\eqref{eq:fittedRAR}. However, RG reproduced the RAR of DMS galaxies with a too large intrinsic scatter ($0.11$~dex vs. $0.057$~dex as found by \mbox{Li et al.~\cite{Li_RAR_2018}} from SPARC galaxies) and with residuals strongly correlated, largely at more than 5$\sigma$, with some galaxy properties dependent on the distance from the galaxies' centre \textcolor{black}{(the radius, the stellar surface density profile, and the gas fraction profile)}, apparently at odds with the results found by Lelli et al.~\cite{Lelli_RAR_2017} \textcolor{black}{for SPARC galaxies. Yet, the RAR of DMS galaxies also presented some, although weaker, correlations between their residuals from Equation~\eqref{eq:fittedRAR} and some galaxy properties. A galaxy sample such as the DMS, made of only 30 galaxies which are also close to being face-on and consequently, with not very accurately measured rotation curves, might not be particularly suitable to investigate the RAR, but a larger sample of edge-on disk galaxies, such as SPARC, should be considered for further analyses. However, further investigation has to be conducted in this sense, since this issue might not depend on the galaxy sample but on the RG theory itself.}  

\section{Dynamics of Elliptical Galaxies}
\label{sec:Elliptical_Galaxies}

In addition to the dynamics of flat systems (disk galaxies), RG has also been tested on the dynamics of spherical systems. To test RG for this class of objects, Cesare et al.~\cite{Cesare_2022} considered a sample of three nearby massive E0 elliptical galaxies, NGC 1407, NGC 4486 (better known ad M87), and NGC 5846, belonging to the \mbox{SLUGGS survey~\cite{Pota_SLUGGS_2013,Brodie_SLUGGS_2014,Forbes_SLUGGS_2017}}, a spectrophotometric survey of more than 4000 extragalactic GCs around 27 early-type galaxies at low redshift. This test was essential to verify whether the boost of the gravitational field could be determined by the gravitational permittivity alone, without any dependence on the refraction of the RG gravitational field lines. The results of this work are also summarised in~\cite{Cesare_2021}.

Since NGC 1407, NGC 4486, and NGC 5846 have a minor-to-major axis ratio $q$ of $0.95$, $0.86$, and $0.92$, and consequently, an ellipticity $\varepsilon = 1 - q$ of $0.05$, $0.14$, and $0.08$, they can be approximated as spherical systems. Cesare et al.~\cite{Cesare_2022} chose these galaxies since the kinematic information of SLUGGS galaxies was measured up to their most external regions ($\gtrsim$10 effective radii), where Newtonian gravity does not work anymore unless DM is present, and the viability of RG could thus be tested. It was possible to measure these extended kinematic profiles thanks to GCs, which are present in these galaxy regions. In general, the measurement of the kinematic information in the external regions of elliptical galaxies can be uniquely performed thanks to the presence of kinematic tracers, such as X-ray emitting gas~\cite{Mathews_and_Brighenti_X-ray_Ellitpicals_2003}, planetary nebulae~\cite{Pulsoni_PNe_Ellipticals_2018}, and GCs~\cite{Pota_SLUGGS_2013}, which settle in the galaxies outskirts. Specifically, the choice of~\cite{Cesare_2022} was driven by the fact that the GC population divides in two subpopulations of blue and red GCs, having different kinematic properties and formation histories. \textcolor{black}{Specifically, the kinematics of the red GCs is usually similar to that of the stars in the host galaxy, maybe due to a similar formation history. Instead, the velocity dispersion of the blue GCs tend to be larger than that of the red GCs~\cite{Pota_SLUGGS_2013}}. In this way, it was possible to constrain RG from two, rather than one, extended kinematic profiles, providing a more robust test for the theory.

\subsection{Mass Model}
\label{sec:Elliptical_Galaxies_Mass_Model}

Cesare et al.~\cite{Cesare_2022} modelled, at the same time and for each E0 galaxy, the root-mean-square (RMS) velocity dispersions of the stars, concentrated within about one effective radius ($R_{\rm e}$; the radius which encloses half the galaxy's luminosity), and of the blue and red GCs, distributed up to the galaxy outskirts. 
In particular, the kinematic profiles of the GCs were $\sim$5, $\sim$48, and $\sim$13 times more extended than the kinematic profiles of the stars in NGC 1407, NGC 4486, and NGC 5846. Cesare et al.~\cite{Cesare_2022} derived the model for the RMS velocity dispersion from the RG gravitational field computed with Equation~\eqref{eq:RGfieldsph} for spherical systems. To model the total mass $M(< r)$ enclosed within the spherical radius $r$ present in Equation~\eqref{eq:RGfieldsph}, Cesare et al.~\cite{Cesare_2022} considered the contribution of (1) the stars, (2), the X-ray emitting gas, and (3) the central super massive black hole (SMBH). The total mass of GCs only contributed $\sim$1\% to the total galaxy mass and it was thus neglected in the computation of the gravitational field. However, their surface and 3D number densities were modelled, since they were needed for the dynamical model (see Section~\ref{sec:Elliptical_Galaxies_Dynamical_Model}).

The quantities entering the dynamical model were differently modelled for the three galaxies. The surface brightness of the stars of NGC 1407, measured in the $B$-band~\cite{Pota_SLUGGS_NGC_1407_2015}, was modelled with a Sérsic profile~\cite{Sersic_1963} of this kind, with parameters from~\cite{Pota_SLUGGS_2013} \textcolor{black}{(Equation~(7) in~\cite{Cesare_2022})}:
\begin{equation}
\label{eq:I_stars_NGC_1407}
I_*(R) = I_{\rm e}\exp\left\{-b_{n_{\rm s}}\left[\left(\frac{R}{R_{\rm e}}\right)^{1/n_{\rm s}} - 1\right]\right\}
\end{equation}
where \textcolor{black}{(Equation (8) in~\cite{Cesare_2022})}
\begin{equation}
    \label{eq:bns}
    b_{n_{\rm s}} \thickapprox 2n_{\rm s} - \frac{1}{3} + \frac{4}{405n_{\rm s}}+\frac{46}{25515n_{\rm s}^2}
\end{equation}
is such that half of the total luminosity of the stars is enclosed within 1 $R_{\rm e}$. The circularised radius $R$ is given for every galaxy by \textcolor{black}{(Equation (6) in~\cite{Cesare_2022})}:
\begin{equation}
    \label{eq:Circularised_Radius}
    R = \sqrt{qx'^2 + \frac{y'^2}{q}},
\end{equation}
where $x'$ and $y'$ are coordinates oriented along the major and minor axes of the galaxies.

Integrating Equation~\eqref{eq:I_stars_NGC_1407} with the Abel integral below yields the 3D luminosity density distribution for the stars of NGC 1407 \textcolor{black}{(Equation (12) in~\cite{Cesare_2022})}:
\begin{equation}
    \label{eq:nu_stars_NGC_1407}
    \nu_*(r) = -\frac{1}{\pi}\int^{+\infty}_r\frac{{\rm d}I_*(R)}{{\rm d}R}\frac{{\rm d}R}{\sqrt{R^2-r^2}},
\end{equation}
and integrating Equation~\eqref{eq:nu_stars_NGC_1407} in spherical coordinates yields its cumulative stellar luminosity profile \textcolor{black}{(Equation (14) in~\cite{Cesare_2022})}: 
\begin{equation}
    \label{eq:L_stars}
    L_*(< r) = 4\pi\int^r_0\nu_*(r')r'^2{\rm d}r'.
\end{equation}

For the surface brightness of the stars of NGC 4486 and NGC 5846, measured in the $r$-band~\cite{Scott_2013}, Cesare et al.~\cite{Cesare_2022} adopted a multi-Gaussian expansion (MGE) model~\cite{Emsellem_1994} in a spherical configuration, with the parameters of~\cite{Scott_2013}  \textcolor{black}{(Equation (10) in~\cite{Cesare_2022})}:
\begin{equation}
    \label{eq:I_stars_NGC_4486_And_NGC_5846}
    I_*(R) = \sum_{k= 1}^{N}\frac{L_k}{2\pi \sigma^2_k}\exp\left(-\frac{R^2}{2\sigma^2_k}\right).
\end{equation}

{The} 
 3D luminosity density distribution of the stars of NGC 4486 and NGC 5846 was given by  \textcolor{black}{(Equation (13) in~\cite{Cesare_2022})}
\begin{equation}
    \label{eq:nu_stars_NGC_4486_And_NGC_5846}
    \nu_*(r) = \sum_{k =1}^{N}\frac{L_k}{(\sqrt{2\pi}\sigma_k)^3}\exp\left(-\frac{r^2}{2\sigma^2_k}\right)
\end{equation}
and their cumulative luminosity stellar profile $L_*(< r)$ was obtained by Equation~\eqref{eq:L_stars}, where $\nu(r')$ is Equation~\eqref{eq:nu_stars_NGC_4486_And_NGC_5846}. 

The cumulative mass profile of the stars in each galaxy, $M_*(< r)$, was obtained by  \textcolor{black}{(Equation (28) in~\cite{Cesare_2022})}:
\begin{equation}
    \label{eq:M_stars}
    M_*(< r) = \Upsilon{L_*(< r)},
\end{equation}
where $L_*(< r)$ is Equation~\eqref{eq:L_stars}, and $\Upsilon$ is the stellar mass-to-light ratio, one of the seven free parameters of the dynamical model.

The 2D number density profiles of the blue and red GCs in NGC 1407 and NGC 5846 were modelled with a Sérsic profile~\cite{Sersic_1963} of this kind  \textcolor{black}{(Equation (15) in~\cite{Cesare_2022})}:
\begin{equation}
\label{eq:N_GCs_NGC_1407_And_NGC_5846}
N_{\rm GC}(R) = N_{\rm e}\exp\left\{-b_{n_s}\left[\left(\frac{R}{R_{\rm e}}\right)^{1/n_s} - 1\right]\right\},
\end{equation}
where Cesare et al.~\cite{Cesare_2022} adopted the parameters of~\cite{Pota_SLUGGS_NGC_1407_2015} for NGC 1407 and estimated the parameters from the data of~\cite{Pota_SLUGGS_2013} with an MCMC for NGC 5846. For NGC 4486, they adopted a different parametrisation  \textcolor{black}{(Equation (16) in~\cite{Cesare_2022})}:
\begin{equation}
    \label{eq:N_GCs_NGC_4486}
    N_{\rm GC}(R) = N_0  \exp\left[-\left(\frac{R}{R_{\rm s}}\right)^\frac{1}{m}\right],
\end{equation}
with the parameters of~\cite{Strader_SLUGGS_NGC_4486_2011}. The 3D number density profiles were obtained by integrating Equations~\eqref{eq:N_GCs_NGC_1407_And_NGC_5846} and~\eqref{eq:N_GCs_NGC_4486} with the Abel integral  \textcolor{black}{(Equation (17) in~\cite{Cesare_2022})}:
\begin{equation}
    \label{eq:nu_GCs}
    \nu_{\rm GC}(r) = -\frac{1}{\pi}\int^{+\infty}_r\frac{{\rm d}N_{\rm GC}(R)}{{\rm d}R}\frac{{\rm d}R}{\sqrt{R^2-r^2}}.
\end{equation}

Cesare et al.~\cite{Cesare_2022} modelled the mass density profile of the gas of NGC 1407 and NGC 5846 using the two-$\beta$ functional form  \textcolor{black}{(Equations (18) and (19) in~\cite{Cesare_2022})}:
\begin{equation}
    \label{eq:rho_gas_NGC_1407_And_NGC_5846}
    \rho_{\rm gas}(r) = \mu{m_{\rm H}}\sqrt{n_{\rm g,1}^2 \left[1 + \left(\frac{r}{R_{\rm c,1}}\right)^2 \right]^{-3\beta_1} + n_{\rm g,2}^2 \left[1 + \left(\frac{r}{R_{\rm c,2}}\right)^2 \right]^{-3\beta_2}},
\end{equation}
with the parameters of~\cite{Zhang_2007} for NGC 1407, and estimating the parameters from the data of~\cite{Paggi_2017} with an MCMC for NGC 5846. In Equation~\eqref{eq:rho_gas_NGC_1407_And_NGC_5846}, $\mu = 0.6$ and $m_{\rm H} = 1.66054 \times 10^{-27}$~kg are the mean molecular weight and the atomic unit mass, respectively. For NGC 4486, they adopted this profile with the parameters of~\cite{Fabricant_1980}  \textcolor{black}{(Equation (20) in~\cite{Cesare_2022})}:
\begin{equation}
    \label{eq:rho_gas_NGC_4486}
    \rho_{\rm gas}(r) = \frac{\rho_0}{(1 + b'r^2 + c'r^4 + d'r^6)^{n'}}.
\end{equation}

The cumulative mass density profile of the gas, $M_{\rm gas}(< r)$, was obtained by integrating in spherical coordinates $\rho_{\rm gas}(r)$  from Equation~\eqref{eq:rho_gas_NGC_1407_And_NGC_5846} for NGC 1407 and NGC 5846, and Equation~\eqref{eq:rho_gas_NGC_4486} for NGC 4486  \textcolor{black}{(Equation (14) in~\cite{Cesare_2022})}:
\begin{equation}
    \label{eq:M_gas}
    M_{\rm gas}(< r) = 4\pi\int^r_0\rho_{\rm gas}(r')r'^2{\rm d}r'.
\end{equation}

The central SMBH was modelled as a point mass with values of $M_\bullet = 4.5^{+0.9}_{-0.4}\times 10^9$~$M_\odot$, $M_\bullet = 6.2^{+0.4}_{-0.5}\times 10^9$~$M_\odot$, and $M_\bullet = 1.1^{+0.1}_{-0.1}\times 10^9$~$M_\odot$ for NGC 1407, NGC 4486, and NGC 5846, obtained from~\cite{Rusli_SMBH_Masses_2013}.

\subsection{Dynamical Model}
\label{sec:Elliptical_Galaxies_Dynamical_Model}

Cesare et al.~\cite{Cesare_2022} \textls[-15]{modelled the RMS velocity dispersion profiles of each kinematic tracer, stars, blue GCs, and red GC, with this solution of the spherical Jeans equations~\cite{Jeans_1915,Mamon_And_Lokas_2005,Cappellari_2008,Pota_SLUGGS_NGC_1407_2015}}  \textcolor{black}{(Equation (23) in~\cite{Cesare_2022})}:
\begin{equation}
    \label{eq:Vrms_Model}
    V_{\rm rms, t}^2(R) = \frac{2}{I_{\rm t}(R)} \int_{R}^{+\infty} K\left(\beta_{\rm t}, \frac{r}{R}\right) \nu_{\rm t}(r)  \frac{{\rm d} \phi}{{\rm d} r} r\text{ }{\rm d}r,
\end{equation}
where (1) ${\rm t} = {*,{\rm B}, {\rm R}}$ indicates the tracer, (2) $I_{\rm t}(R)$ is either the surface brightness of the stars (Equations~\eqref{eq:I_stars_NGC_1407} or~\eqref{eq:I_stars_NGC_4486_And_NGC_5846}) or the 2D number density of the GCs (Equations~\eqref{eq:N_GCs_NGC_1407_And_NGC_5846} or~\eqref{eq:N_GCs_NGC_4486}), (3) $\nu_{\rm t}(r)$ is either the 3D luminosity density of the stars (Equations~\eqref{eq:nu_stars_NGC_1407} or~\eqref{eq:nu_stars_NGC_4486_And_NGC_5846}) or the 3D number density of the GCs (Equation~\eqref{eq:nu_GCs}), (4) $\phi$ is the gravitational potential and, thus, ${\rm d}\phi/{\rm d}r$ is the gravitational field (Equation~\eqref{eq:RGfieldsph}), (5) $R$ is the 2D radius (Equation~\eqref{eq:Circularised_Radius}), (6) $r$ is the 3D spherical radius, (7) $\beta_{\rm t} = 1 - \sigma_\theta^2/\sigma_r^2$ is the orbital anisotropy parameter, where $\sigma_\theta$ and $\sigma_r$ are the tangential and radial velocity dispersions, and (8) $K$ is the kernel \textcolor{black}{(Equation~(A16) in~\cite{Mamon_And_Lokas_2005} and Equation (24) in~\cite{Cesare_2022})}:
\begin{equation}
    \label{eq:Kernel_beta}
	\begin{split}
	K\left(\beta_{\rm t}, \frac{r}{R}\right) = &\frac{1}{2}\left(\frac{r}{R}\right)^{2\beta_{\rm t}-1}\bigg[\left(\frac{3}{2}-\beta_{\rm t}\right)\sqrt{\pi}\frac{\Gamma(\beta_{\rm t}-\frac{1}{2})}{\Gamma(\beta_{\rm t})}\\
	&+\beta_{\rm t} B_{\frac{R^2}{r^2}}\left(\beta_{\rm t} + \frac{1}{2}, \frac{1}{2}\right) - B_{\frac{R^2}{r^2}}\left(\beta_{\rm t} - \frac{1}{2}, \frac{1}{2}\right) \bigg], \\
	\end{split}
\end{equation}
where $\Gamma(z)$ is the Euler gamma function, and $B_x(a,b)$ is the incomplete beta function. The orbital anisotropy parameters $\beta_*$, $\beta_{\rm B}$, and $\beta_{\rm R}$ are three of the seven free parameters of the dynamical model, and they are assumed to be constant with the radius~\cite{Thomas_2014,Rantala_2019}. 

The gravitational field ${\rm d}\phi/{\rm d}r$ in Equation~\eqref{eq:Vrms_Model} is given by the RG field in a spherical configuration (Equation~\eqref{eq:RGfieldsph}), and thus, Equation~\eqref{eq:Vrms_Model} transforms into  \textcolor{black}{(Equation (25) in~\cite{Cesare_2022})}:
\begin{equation}
    \label{eq:Vrms_Model_RG}
    V_{\rm rms, t}^2(R) = \frac{2G}{I_{\rm t}(R)} \int_{R}^{+\infty} K\left(\beta_{\rm t}, \frac{r}{R}\right) \nu_{\rm t}(r) \frac{M(< r)}{\epsilon(\rho)} \frac{{\rm d}r}{r}.
\end{equation}

The cumulative mass profile $M(< r)$ in Equation~\eqref{eq:Vrms_Model_RG} is given by the sum of the cumulative mass profiles of the stars $M_*(< r)$ (Equation~\eqref{eq:M_stars}), the gas $M_{\rm gas}(< r)$ (Equation~\eqref{eq:M_gas}), and the SMBH \textcolor{black}{(end of Section~\ref{sec:Elliptical_Galaxies_Mass_Model}):} 
\begin{equation}
    \label{eq:M_total}
    M(< r) = M_*(< r) + M_{\rm gas}(< r) + M_\bullet,
\end{equation}
\textcolor{black}{taken from Equation (26) in~\cite{Cesare_2022}.}

This dynamical model has seven free parameters. Four parameters are common to the three tracers, the stellar mass-to-light ratio $\Upsilon$, and the three RG parameters, $\epsilon_0$, $Q$, and $\rho_{\rm c}$, entering the gravitational permittivity $\epsilon(\rho)$ in Equation~\eqref{eq:Vrms_Model_RG}, which generate the RG gravitational potential to which every kinematic tracer is subject. Three parameters are specific to each tracer, and they are the orbital anisotropy parameters $\beta_*$, $\beta_{\rm B}$, and $\beta_{\rm R}$.

To explore the parameter space, Cesare et al.~\cite{Cesare_2022} adopted a MCMC method based on a Metropolis--Hastings acceptance criterion. They assumed a uniform prior on every free parameter. For $\Upsilon$, the uniform prior was based on the SPS models of Humphrey et al.~\cite{Humphrey_2006} and Zhang et al.~\cite{Zhang_2007} for NGC 1407, and of Bell et al.~\cite{Bell_2003} and Zibetti et al.~\cite{Zibetti_2009} for NGC 4486 and NGC 5846. The space of the three RG free parameters was explored in the ranges of $(0.00,1.00]$, $[0.01,2.00]$, and $[-27.00,-23.00]$, for $\epsilon_0$, $Q$, and $\log_{10}[\rho_{\rm c}({\rm g}/{\rm cm}^3)]$, respectively. At last, they explored the parameter space of the parameters $\mathcal{B}_*=-\log_{10}(1-\beta_*)$, $\mathcal{B}_{\rm B}=-\log_{10}(1-\beta_{\rm B})$,  $\mathcal{B}_{\rm R}=-\log_{10}(1-\beta_{\rm R})$ in the uniform range of $[-1.5,1.0]$, which spans from very tangential to very radial orbits.  

The model given by Equation~\eqref{eq:Vrms_Model_RG} was compared with the kinematic measurements of~\cite{Pota_SLUGGS_NGC_1407_2015} for NGC 1407, and of the ATLAS$^{\rm 3D}$ survey~\cite{Cappellari_ATLAS3D_2011} for NGC 4486 and NGC 5846, to constrain the free parameters.

\subsection{Results}
\label{sec:Elliptical_Galaxies_Results}

\begin{figure}[H]
    		\includegraphics[scale=0.80]%
    		{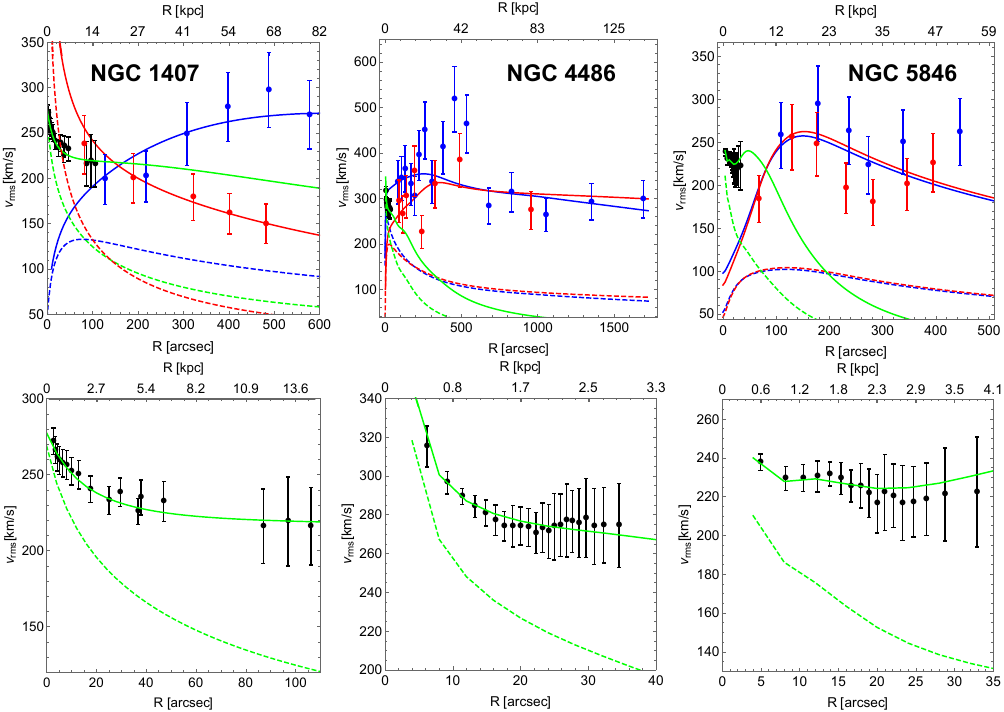}
    		\caption{Root-mean-square velocity dispersion profiles computed with RG with the parameters resulting from the MCMC analysis of~\cite{Cesare_2022}, considering the stars, blue GCs, and red GCs' kinematic profiles at the same time for each elliptical E0 galaxy. (\textit{\bf Top panels}) the green, blue, and red solid lines represent the RG models for the stars, blue GCs, and red GCs, respectively. The dashed lines with the same colours are computed with the same $\Upsilon$ and $\beta$ parameters as the solid lines but with Newtonian gravity without DM. (\textit{\bf Bottom panels}) zoom-in of the upper panels for the stars alone. Green solid and dashed lines have the same meaning as in the top panels. Both in the top and in the bottom panels, the black, blue, and red dots with error bars show the kinematic data for the stars, blue GCs, and red GCs, respectively. The figure is reproduced from Figure 8 in~\cite{Cesare_2022}. Credit: Cesare V., Diaferio A., and Matsakos T., A\&A, 657, A133, 2022, reproduced with permission
$\copyright$ ESO.}
    		\label{fig:V_RMS_RG_Elliptical_Galaxies}
\end{figure}

The RG model~\eqref{eq:Vrms_Model_RG} properly interpolated the kinematic profiles of the three tracers in each E0 galaxy with sensible parameters (Figure~\ref{fig:V_RMS_RG_Elliptical_Galaxies}). 

The resulting mass-to-light ratios were consistent with the corresponding SPS models, and the orbital anisotropy parameters of NGC 1407 were in agreement with the literature~\cite{Pota_SLUGGS_NGC_1407_2015}. The RG parameters were consistent among the three galaxies within 2$\sigma$, which supported their universality (orange squares with error bars in Figure~\ref{fig:RG_Parameters_Disk_Galaxies_Elliptical_Galaxies}). The mean of the values resulting from the three single galaxies were $\epsilon_{0,{\rm Mean, Ell}} = 0.089^{+0.038}_{-0.035}$, $Q_{\rm Mean, Ell} = 0.47^{+0.29}_{-0.21}$, and $\log_{10}[\rho_{\rm c}({\rm g}/{\rm cm}^3)]_{\rm Mean, Ell} = -24.25^{+0.28}_{-0.20}$. To better check their universality, Cesare et al.~\cite{Cesare_2022} compared these values with the mean and the unique combinations of RG parameters resulting from the DMS galaxies, $\epsilon_{0,{\rm Mean,DMS}} = 0.56 \pm 0.16$, $Q_{\rm Mean,DMS} = 0.92 \pm 0.71$, $\log_{10}[\rho_{\rm c}({\rm g}/{\rm cm}^3))]_{\rm Mean,DMS} = -25.30 \pm 1.22$ (purple squares with error bars in Figure~\ref{fig:RG_Parameters_Disk_Galaxies_Elliptical_Galaxies}), and $\epsilon_{0,{\rm Unique,DMS}} = 0.661^{+0.007}_{-0.007}$, $Q_{\rm Unique,DMS} = 1.79^{+0.14}_{-0.26}$, $\log_{10}[\rho_{\rm c}({\rm g}/{\rm cm}^3))]_{\rm Unique,DMS} = -24.54^{+0.08}_{-0.07}$ (green dots with blue shaded posterior distributions in Figure~\ref{fig:RG_Parameters_Disk_Galaxies_Elliptical_Galaxies}). The RG parameters derived from the elliptical galaxies were consistent with the mean combination obtained in the DMS: $Q$ and $\log_{10}[\rho_{\rm c}({\rm g}/{\rm cm}^3))]$ within 1$\sigma$ and $\epsilon_0$ within 3$\sigma$. The parameters $Q$ and $\log_{10}[\rho_{\rm c}({\rm g}/{\rm cm}^3))]$ from the E0 galaxies were also in agreement with the unique values from the DMS, within $3.4$$\sigma$ and $1.4$$\sigma$, respectively. Instead, the $\epsilon_0$ parameter presented a $14.8$$\sigma$ tension.

The reasons for the $\epsilon_0$ tension can be multiple. A first possible reason could be due to the approximate procedure with which the unique combination of RG parameters in the DMS was derived. A limited freedom was given to this unique combination of parameters to vary, since the mass parameters, $\Upsilon$ and $h_z$, were constrained to the values derived from a previous analysis and this led to a very small width of the posterior distribution of $\epsilon_0$ (see the light blue shaded area and yellow, red, and black contours in the bottom-left panel of Figure~\ref{fig:RG_Parameters_Disk_Galaxies_Elliptical_Galaxies}). As mentioned in Section~\ref{sec:Disk_Galaxies_Results}, a more complete analysis could be performed by exploring the 63-dimensional parameter space of the $2 \times 30$ mass parameters and the three RG parameters in the DMS. 

If, instead, the posterior distribution of $\epsilon_0$ is not particularly underestimated, other reasons could \textcolor{black}{explain} the $\epsilon_0$ discrepancy, such as a too approximated model for the elliptical galaxies, a wrong functional form for the gravitational permittivity $\epsilon(\rho)$, or a fundamental problem with RG. One of the approximations introduced in the analysis of elliptical galaxies was to consider these systems as isolated, whereas these kinds of galaxies typically live in dense environments. Specifically, NGC 1407 and NGC 5846 settle within galaxy groups and NGC 4486 is the central galaxy of the Virgo cluster. Differently from Newtonian gravity, the RG gravitational field of a system, as a MOND gravitational field, depends on the gravitational field produced by the environment where this system settles. Therefore, this environmental effect might explain, in part or entirely, the inconsistency between the $\epsilon_0$s. Another approximation adopted in modelling the elliptical galaxies was to neglect their net rotation, although these three galaxies are slow rotators~\cite{Cesare_2022}. \textcolor{black}{A caveat should be given since these elliptical galaxies were approximated as spherical systems due to their low minor-to-major axis ratio $q$. However, they could be significantly flattened oblate or prolate spheroidal systems seen along the symmetry axis.} Moreover, or as an alternative, the functional form of the gravitational permittivity (Equation~\eqref{eq:epsilon}) might not be suitable, in particular in the low-density regime, and a more precise answer in this sense might be given by the weak field limit (WFL) of the covariant formulation of RG~\cite{Sanna_CRG_2023}. If all these reasons are not able to alleviate or cancel the $\epsilon_0$ tension, this might be due to a fault in the theory that needs to be fixed. To conclude, a remarkable fact is that the best agreement between elliptical and disk galaxies was obtained for the $\rho_{\rm c}$ parameter, which sets the density scale where the transition between the Newtonian and the RG regimes occurs, as $a_0$ in MOND.

\section{Dynamics of Galaxy Clusters}
\label{sec:Galaxy_Clusters}

The encouraging results obtained for the dynamics of disk and elliptical galaxies suggest that RG is able to properly describe the dynamics on a galaxy scale. Therefore, a class of tests at a larger scale need to be performed for a more complete investigation of the~theory. 

\begin{figure}[H]
 \hspace{-13pt} 		\includegraphics[scale=0.60]%
    	{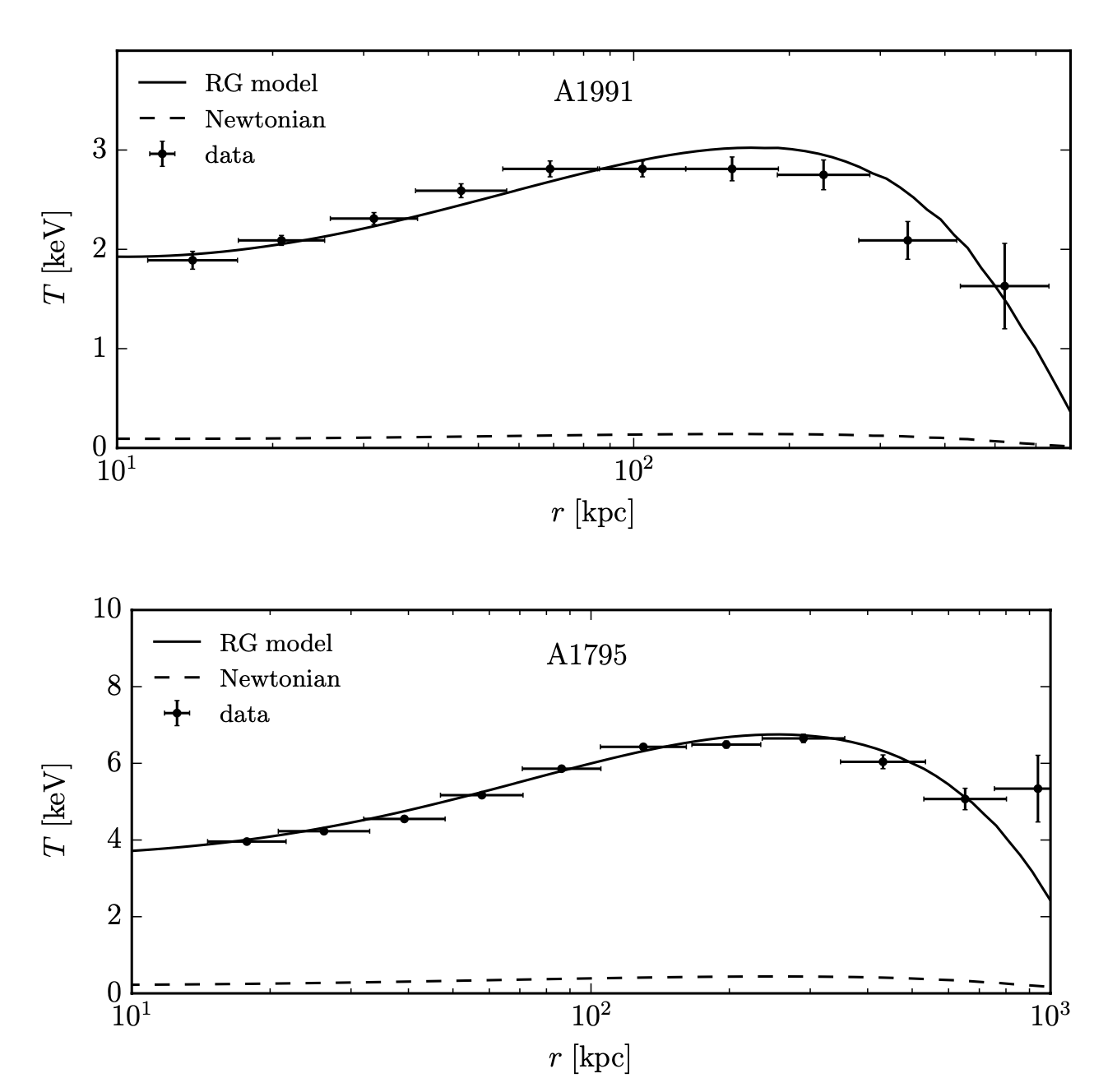}
    		\caption{\textcolor{black}{Emission-weighted projected} temperature radial profiles of the hot, X-ray-emitting gas in the low-redshift and relaxed galaxy clusters A1991 (\textbf{top panel}) and A1795 (\textbf{bottom panel}). The solid and dashed lines are the RG and Newtonian models, respectively, and the dots with error bars are the measurements taken with the Chandra satellite from~\cite{Vikhlinin_Chandra_2006}. \textcolor{black}{The Newtonian expectations were calculated with the same mass model as in RG without considering the presence of DM.} A Hubble constant of $H_0 = 71$~km~s$^{-1}$~Mpc$^{-1}$ was adopted. The figure is reproduced from Figure 14 in~\cite{Matsakos_and_Diaferio_RG_2016}.}
    		\label{fig:RG_Galaxy_Clusters}
\end{figure}

Some preliminary studies have already been performed by Matsakos and Diaferio~\cite{Matsakos_and_Diaferio_RG_2016} to study the viability of RG in modelling the hot X-ray-emitting gas temperature radial profile of galaxy clusters. Matsakos and Diaferio~\cite{Matsakos_and_Diaferio_RG_2016} modelled the gas temperature radial profiles of two low-redshift and relaxed galaxy clusters, A1991 and A1795, with two different gas temperatures (Figure~\ref{fig:RG_Galaxy_Clusters}). Specifically, the two clusters have a spectroscopic gas temperature, averaged between 70 kpc and $r_{500}$, of $2.83$ keV (A1991) and $9.68$ keV (A1795), where $r_{500}$ is the distance from the centre of the cluster where the average density is 500~times larger than the critical density of the Universe \textcolor{black}{if DM is assumed to be present~\cite{Vikhlinin_Chandra_2006}}. These data are taken with the Chandra satellite~\cite{Vikhlinin_Chandra_2006}. The model that Matsakos and Diaferio~\cite{Matsakos_and_Diaferio_RG_2016} adopted presented different assumptions. They assumed dynamical equilibrium at every distance from the cluster centre and a spherical configuration for the density distribution, deriving the equation of hydrostatic equilibrium from the RG gravitational field obtained, in turn, from Equation~\eqref{eq:RGfieldsph}. They also assumed the validity of the ideal-gas equation of state and the smooth step function given by Equation~\eqref{eq:epsilon} for the gravitational permittivity with RG parameters fixed a priori and not left free to vary. In particular, they set the RG parameters to $\epsilon_0 = 0.045$ and $0.065$ for A1991 and A1795, respectively, $Q = 2$, and $\rho_{\rm c} = 10^{-24}$~g~cm$^{-3}$. At last, they assumed in A1991 the presence of a stellar component of mass $5 \times 10^{10}$~M$_\odot$ up to a distance from the cluster centre of $R < 10$ kpc. This model resulted in an agreement with the Chandra data of the two galaxy clusters. \textcolor{black}{Figure~\ref{fig:RG_Galaxy_Clusters} shows the RG models as black solid lines and the Newtonian expectations calculated with the same mass models as in RG, without considering any DM presence, as black dashed lines.}
   \vspace{-11pt}

\textls[-15]{This model is certainly simplistic and involves a too small sample of galaxy clusters. Therefore, it certainly needs to be extended to understand if RG is able to correctly describe the gas temperature radial profiles of galaxy clusters, and further tests have to be performed to investigate the capacity of RG to model the dynamics of these systems. An important test that has to be performed is to repeat the above analysis by constraining the RG parameters from the temperature data, and not by fixing them a priori, to verify if the obtained parameters are in agreement with the values obtained on a galaxy scale, to better test their universality.}

Another aspect of RG on a cluster scale that Matsakos and Diaferio~\cite{Matsakos_and_Diaferio_RG_2016} began to explore is the gravitational interaction between galaxies in galaxy groups and clusters. These interactions are not simple to handle in RG, since the field lines in each galaxy suffer from refraction in low-density regions \textcolor{black}{distant from their centres}. A first-order analysis could be performed by assuming that the gravitational field at large distances from the centre of each galaxy decreases proportional to $R^{-1}$ and that the only difference with respect to Newtonian \textcolor{black}{gravitational field} is that it is enhanced by $\epsilon_0^{-1}$. With these assumptions, \textcolor{black}{the BTFR (Equation~\eqref{eq:BTFR}) can be extended to galaxy groups and clusters. In particular,} the following expression for the BTFR would hold for galaxy groups and clusters \textcolor{black}{(\mbox{Equation (6.1)} in~\cite{Matsakos_and_Diaferio_RG_2016})}:
\begin{equation}
    \label{eq:BTFR_extended_galaxy_groups_and_clusters}
    v_{\rm f}^4 = Gbf\frac{kM}{\epsilon_0},
\end{equation}
\textcolor{black}{where} $k$ is the number of galaxies, $b$ is the BTFR normalisation, and $f < 1$ is a geometric factor that describes that, accounting for the anisotropic geometry of the RG field in disk galaxies, only a fraction of the $k$ galaxies in the clusters will cross the plane of a specific disk galaxy where the field is $\propto R^{-1}$. \textcolor{black}{In Equation~\eqref{eq:BTFR_extended_galaxy_groups_and_clusters}, $v_{\rm f}$ represents the circular velocity of the galaxies in a specific galaxy group or cluster with total baryonic mass $M$ (mainly made by the intracluster medium (ICM)), whereas in Equation~\eqref{eq:BTFR}, $v_{\rm f}$ represents the circular velocity of the stars in a specific galaxy with total baryonic mass $M$.} 

\begin{figure}[H]
    		\includegraphics[scale=0.430]%
   	{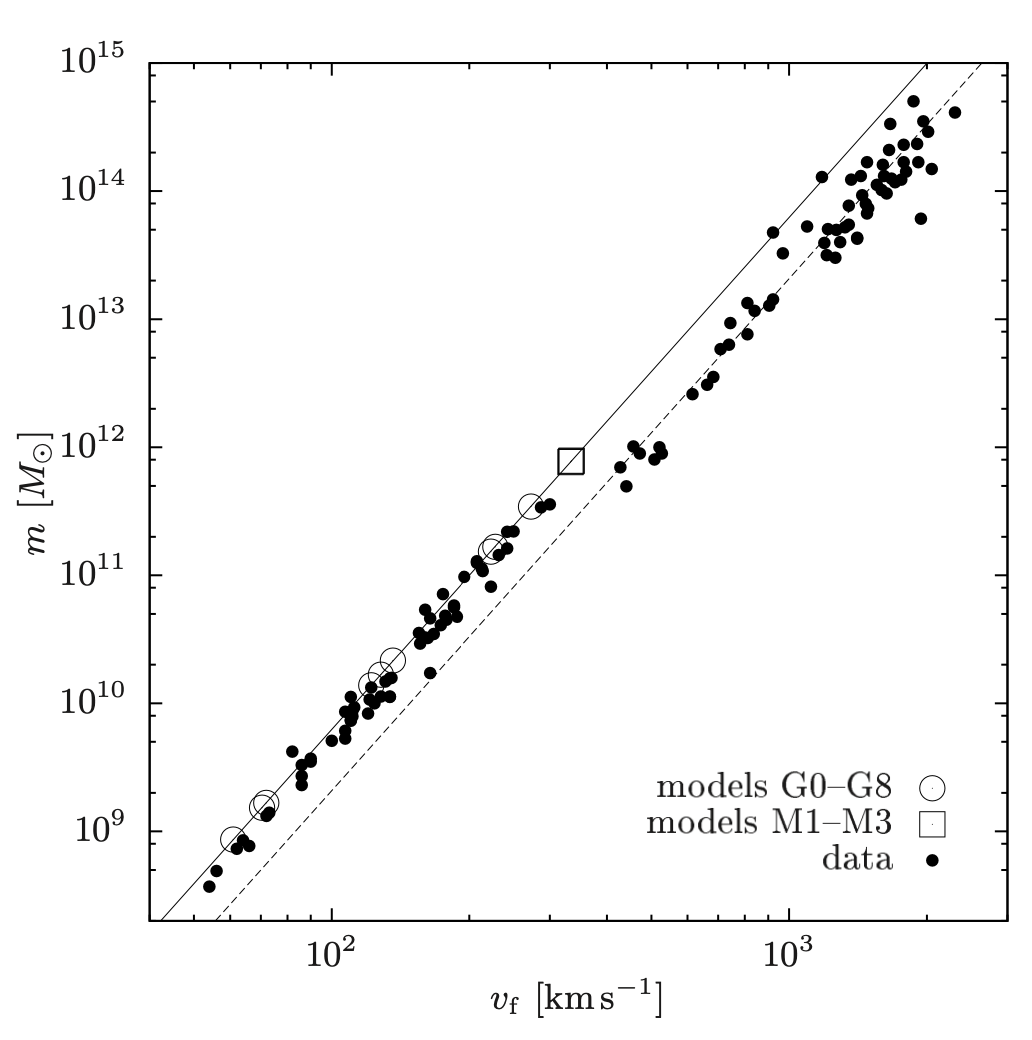}
    	\caption{\textls[-20]{The BTFR up to mass and velocity scales to include galaxy groups and clusters, besides galax-\linebreak  ies. The black solid line is the BTFR given by Equation~\eqref{eq:BTFR}, and the black dashed line is the BTFR given by  }} \label{fig:BTFR_Galaxies_Galaxy_Groups_And_Galaxy_Clusters_RG}
\end{figure}

\textcolor{black}{Figure~\ref{fig:BTFR_Galaxies_Galaxy_Groups_And_Galaxy_Clusters_RG} represents the BTFR extended to galaxy groups and clusters (Equation~\eqref{eq:BTFR_extended_galaxy_groups_and_clusters}, black dashed line) for $f/\epsilon_0 = 1.3$ and the BTFR of galaxies (Equation~\eqref{eq:BTFR}, black solid line). The black dots in the bottom-left part of the plot represent data points for galaxies taken from~\cite{McGaugh_2005}, having a mass range of $M \sim [10^9,10^{12}]$~M$_\odot$ and the black dots in the top-right part of the plot represent data points for galaxy groups and clusters taken from~\cite{Sanders_2003,McGaugh_2015}, having a mass range of $M \sim [10^{12},10^{15}]$~M$_\odot$. We can see that the two sets of data points distribute around the corresponding BTFR model. The open circles and squares represent simulated galaxies and they distribute around Equation~\eqref{eq:BTFR} (black solid line). The observed normalisation of the BTFR extended to galaxy groups and clusters is smaller than the normalisation of the BTFR of galaxies but its slope is the same as for Equation~\eqref{eq:BTFR}, consistent with the approximate RG expectations.} The comparison of this extended BTFR with the observed one for galaxy groups and clusters would represent another test for RG and a method to set a lower bound for $\rho_{\rm c}$ and to constrain $\epsilon_0$.
\vspace{-12pt}
{ \captionof*{figure}{Equation~\eqref{eq:BTFR_extended_galaxy_groups_and_clusters} for $f/\epsilon_0 = 1.3$. The open circles and squares represent simulated galaxies~\cite{Matsakos_and_Diaferio_RG_2016}. The black dots that concentrate in the bottom-left part of the plot are observational data from~\cite{McGaugh_2005} and refer to galaxies. The black dots that concentrate in the top-right part of the plot are observational data from~\cite{Sanders_2003,McGaugh_2015} and refer to galaxy groups and clusters. The two set of measurement points concentrate around the corresponding BTFR model. The figure is reproduced from Figure 17 in~\cite{Matsakos_and_Diaferio_RG_2016}.}}
\vspace{12pt}

\section{Covariant Refracted Gravity}
\label{sec:Covariant_RG}

A covariant version of RG was recently formulated by Sanna et al.~\cite{Sanna_CRG_2023}. The fact that in RG, the transition between the Newtonian and the modified gravity regimes is regulated by a scalar quantity, the mass density $\rho$, rather than by a vector quantity, such as the acceleration as in MOND, made the building of a covariant extension of RG less challenging than in MOND (e.g.,~\cite{Bekenstein_2004}). Some relativistic extensions of MOND, such as tensor vector scalar gravity (TeVeS)~\cite{Bekenstein_2004}, were formulated while presenting some problems~\cite{Skordis_2009,Bekenstein_2011}, even if more recent results might look more promising~\cite{Hernandez_2019,Skordis_and_Zlosnik_2021}. 

Covariant refracted gravity (CRG) is formulated as a scalar tensor theory with the presence of a single scalar field, $\varphi$, nonminimally coupled to the metric, which accounts for the phenomenologies of both DM on a galaxy scale and DE on larger scales, i.e., explaining the accelerated cosmic expansion. This peculiar feature of a unified dark sector is shared by a restricted class of modified theories of gravity (e.g.,~\cite{Kunz_UDS_2009a,Kunz_UDS_2009b,Makler_UDS_2003,Bento_UDS_2003,Scherrer_UDS_2004,Bruni_UDS_2013,Cadoni_UDS_2018,Sotiriou_and_Faraoni_UDS_2010,Sebastiani_and_Vagnozzi_UDS_2016,Berezhiani_UDS_2017,Ferreira_UDS_2019,Arbey_and_Coupechoux_2021,Campigotto_UDS_2019}) and is suggested by the $a_0 \sim H_0 \sim \sqrt{\Lambda}$ observed intriguing coincidence~\cite{Sanna_CRG_2023} (see Section~\ref{sec:Introduction}).

The general action of scalar--tensor theories is \textcolor{black}{(Equation (3) in~\cite{Sanna_CRG_2023})}:
\begin{equation}
    \label{eq:ST_action}
    \mathcal{S} = \frac{1}{16\pi G} \int {\rm d}^4x \sqrt{g}\left[\varphi R + \frac{\mathcal{W}(\varphi)}{\varphi}\nabla^\alpha\varphi\nabla_\alpha\varphi + 2 \mathcal{V}(\varphi)\right] + \int {\rm d}^4x \sqrt{g} \mathcal{L}_{\rm m}(g_{\mu\nu},\psi_{\rm m}),
\end{equation}
where the functional forms of the self-interaction potential $\mathcal{V}(\varphi)$ and the general differentiable function $\mathcal{W}(\varphi)$ of the scalar field $\varphi$ define a specific scalar--tensor theory. For CRG, $\mathcal{V}(\varphi)$ and $\mathcal{W}(\varphi)$ are \textcolor{black}{(Equation (7) in~\cite{Sanna_CRG_2023})}:
\begin{equation}
    \label{eq:V_varphi_CRG}
    \mathcal{V}(\varphi) = -\Xi\varphi,
\end{equation}
and \textcolor{black}{(Equation (6) in~\cite{Sanna_CRG_2023})}
\begin{equation}
    \label{eq:W_varphi_CRG}
    \mathcal{W}(\varphi) = -1,
\end{equation}
where $\Xi$ is a constant.
Replacing Equations~\eqref{eq:V_varphi_CRG} and~\eqref{eq:W_varphi_CRG} in Equation~\eqref{eq:ST_action}, the CRG action becomes:
\begin{equation}
    \label{eq:ST_action_CRG}
    \mathcal{S}_{\rm CRG} = \frac{1}{16\pi G} \int {\rm d}^4x \sqrt{g}\left[\varphi R - \frac{\nabla^\alpha\varphi\nabla_\alpha\varphi}{\varphi} - 2\Xi\varphi\right] + \int {\rm d}^4x \sqrt{g} \mathcal{L}_{\rm m}(g_{\mu\nu},\psi_{\rm m}).
\end{equation}

In the WFL of the theory, the scalar field $\varphi$ becomes twice the gravitational permittivity $\epsilon(\rho)$. Indeed, the WFL of CRG yields the original classical formulation of RG \textcolor{black}{(Equation (24) in~\cite{Sanna_CRG_2023})}:
\begin{equation}
    \label{eq:WFL_CRG}
    \nabla\cdot (\varphi\nabla\phi) \simeq 8\pi G\rho,
\end{equation}
which is the RG Poisson equation (Equation~\eqref{eq:RG_Poisson_Equation}) if $\varphi = 2\epsilon(\rho)$. This result confirms that the scalar field mimics the phenomenology of DM on a galaxy scale. 

The fact that in RG, the modification of the law of gravity depends on a density rather than on an acceleration scale might be less intuitive, since the acceleration scale $a_0$ emerges from several pieces of evidence on a galaxy scale, such as the BTFR, the MDAR, and the RAR (see Section~\ref{sec:Introduction}). Yet, the acceleration scale $a_0$ emerges from the WFL of CRG. This can be seen by calculating the CRG gravitational field far from a spherical source with density $\rho_{\rm s}(r)$, monotonically decreasing with $r$, settling in a homogeneous background with constant density $\rho_{\rm bg}$. Sanna et al.~\cite{Sanna_CRG_2023} found that, far from this source, the transition from Newtonian to RG regimes occurs when the acceleration ${\rm d}\phi/{\rm d}r$ goes below the acceleration scale \textcolor{black}{(Equation (28) in~\cite{Sanna_CRG_2023})}:
\begin{equation}
     \label{eq:acceleration_scale_CRG}    
     a_\Xi = \sqrt{2\Xi - \frac{8\pi G\rho}{\varphi}},
\end{equation}
where $\rho(r) = \rho_{\rm s}(r) + \rho_{\rm bg}$. This recalls the MOND theory, where the acceleration scale $a_0$ regulates the gravity behaviour. At large distances from the spherical source, the following limit holds \textcolor{black}{(Section 3.4.1  in~\cite{Sanna_CRG_2023})}:
\begin{equation}
    \label{eq:CRG_limit_source_large_dist}
    2\Xi \gg 8\pi G\rho/\varphi,
\end{equation}
and therefore, Equation~\eqref{eq:acceleration_scale_CRG} reduces to \textcolor{black}{(Section  3.4.1  in~\cite{Sanna_CRG_2023})}:
\begin{equation}
    \label{eq:acceleration_scale_CRG_limit_source_large_dist}
    a_\Xi \sim \sqrt{2\Xi}.
\end{equation}

Since $\Xi \sim \Lambda$, as found by independent calculations~\cite{Sanna_CRG_2023} explained below, Equation~\eqref{eq:acceleration_scale_CRG_limit_source_large_dist} implies that $a_\Xi \sim 10^{-10}$~m~s$^{-2}$, which coincides with the MOND acceleration scale $a_0$. 

Sanna et al.~\cite{Sanna_CRG_2023} also derived with CRG the modified Friedmann equations for a flat, expanding, homogeneous, and isotropic Universe described with the Friedmann--Lema$\hat{i}$tre--Robertson--Walker (FLRW) metric, where the Universe content is modelled as a perfect fluid. In these modified Friedmann equations, the term \textcolor{black}{(Equation (41) in~\cite{Sanna_CRG_2023})}
\begin{equation}
    \label{eq:Omega_Xi_Term_Modified_Friedmann_Equations_CRG}
    \Omega_\Xi = \frac{\Xi}{3H^2}
\end{equation}
appears, where $H(t)$ is the Hubble parameter. This term is analogous to the density parameter related to the cosmological constant $\Lambda$ in $\Lambda$CDM \textcolor{black}{(Section  4.2  in~\cite{Sanna_CRG_2023})}:
\begin{equation}   
\label{eq:Omega_Lambda_Term_LambdaCDM}
    \Omega_\Lambda = \frac{\Lambda}{3H^2},
\end{equation}
which indicates that $\Xi \sim \Lambda$, i.e, that $\Xi$ plays the role of $\Lambda$ in $\Lambda$CDM, accounting for the accelerated expansion of the Universe. At the present time, Equations~\eqref{eq:Omega_Xi_Term_Modified_Friedmann_Equations_CRG} and~\eqref{eq:Omega_Lambda_Term_LambdaCDM} become:
\begin{equation}   \label{eq:Omega_Xi_Term_Modified_Friedmann_Equations_CRG_Present_Time}
    \Omega_{\Xi,0} = \frac{\Xi}{3H_0^2},
\end{equation}
and
\begin{equation}   
\label{eq:Omega_Lambda_Term_LambdaCDM_Present_Time}
    \Omega_{\Lambda,0} = \frac{\Lambda}{3H_0^2},
\end{equation}
where $H_0$ is the Hubble parameter at the present time. We have seen before that $\Xi$ sets the value of the acceleration scale $a_\Xi$ (Equation~\eqref{eq:acceleration_scale_CRG}), which defines the transition between Newtonian and RG regimes far from a spherical source, playing the role of $a_0$ in MOND and thus accounting for DM phenomenology. Since $\Xi \sim \Lambda$, $\Xi$ also accounts for DE phenomenology, providing a unification of the two dark sectors. Moreover, inserting $\Xi$ in Equation~\eqref{eq:acceleration_scale_CRG}, the observed relation $a_0 \sim \sqrt{\Lambda}$ naturally emerges in CRG.

\begin{figure}[H]
    		\includegraphics[scale=1.50]%
   	{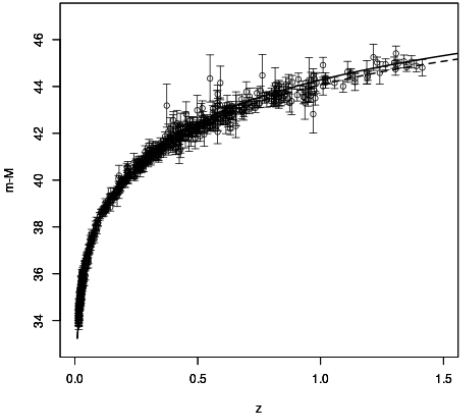}
    	\caption{Hubble diagram of \textcolor{black}{SNe Ia} modelled with CRG. The dashed and solid lines are the CRG$-$ and CRG$+$ models, respectively. For the CRG$-$ and CRG$+$ curves, the $H_0 = 67.7$~km~s$^{-1}$~Mpc$^{-1}$, $\Omega_0 = 0.31$, and $\Omega_{\Xi,0}= 0.65$ parameters were adopted. The open circles with error bars are the data from the Supernova Cosmology Project Union $2.1$ Compilation~\cite{Suzuki_2012}. The figure is reproduced from Figure E.2 in~\cite{Sanna_CRG_2023}.}	\label{fig:SNaeIa_Hubble_Diagram_CRG}
\end{figure}

By rearranging the modified Friedmann equations in CRG, Sanna et al.~\cite{Sanna_CRG_2023} found two solutions for the Hubble parameter $H(t)$, referred to as CRG$-$ and CRG$+$, and properly integrating $H(t)$, they found the corresponding two solutions for the scale factor $a(t)$. Properly combining CRG$-$ and CRG$+$, the bound $\Xi > 0$ was derived. From the derived $a(t)$, the luminosity distance $D_{\rm L}$ could be calculated and constrained from the observed Hubble diagram of \textcolor{black}{SNe Ia} at high redshift $z$ of the Supernova Cosmology Project Union $2.1$ Compilation~\cite{Suzuki_2012} \textcolor{black}{(open circles with error bars in Figure~\ref{fig:SNaeIa_Hubble_Diagram_CRG}).} Comparing the CRG model with the data, the cosmological parameters could be derived to verify if the tensions observed in $\Lambda$CDM could be either reduced or cancelled, which would provide a fundamental test for CRG. By comparing the distance modulus $\mu = m - M$, where $m$ and $M$ are the apparent and absolute magnitudes, respectively, derived from the luminosity distance $D_{\rm L}$ computed in CRG for the solutions CRG$-$ and CRG$+$ \textcolor{black}{(dashed and solid lines in Figure~\ref{fig:SNaeIa_Hubble_Diagram_CRG})} with the $(z,\mu)$ data of the high-redshift \textcolor{black}{SNe Ia} from~\cite{Suzuki_2012}, Sanna et al.~\cite{Sanna_CRG_2023} found that $\Omega_{\Xi,0} = 0.650^{+0.005}_{-0.650}$ for CRG$-$, and $\Omega_{\Xi,0} = 0.650^{+0.005}_{-0.085}$ for CRG$+$, assuming the values $H_0 = 67.7$ km s$^{-1}$ Mpc$^{-1}$ and $\Omega_0 = 0.31$ for the Hubble constant and the ratio between the densities of the baryonic matter and the critical density of the Universe at the present epoch, respectively~\cite{Planck_2020}. Replacing the values of $\Omega_{\Xi,0} = 0.65$ and $H_0 = 67.7$ km s$^{-1}$ Mpc$^{-1}$ in Equation~\eqref{eq:Omega_Lambda_Term_LambdaCDM_Present_Time} implies again that $\Xi \sim \Lambda$, as independently determined before.

Sanna et al.~\cite{Sanna_CRG_2023} derived the equation of state of the effective DE in CRG, $w_{\rm DE} = p_{\rm DE}/\rho_{\rm DE}$, where $p_{\rm DE}$ and $\rho_{\rm DE}$ are the pressure and the density of the effective DE, by properly rearranging the CRG's Friedmann equations and by comparing them with the Friedmann equations of a general scalar--tensor theory with a nonminimal coupling between the scalar field and the metric. The $w_{\rm DE}$ parameter depends on redshift $z$. At the present time, $z = 0$, the $w_{\rm DE}$ parameter becomes \textcolor{black}{(Equation (73) in~\cite{Sanna_CRG_2023})}:
\begin{equation}
    \label{eq:w_DE_CRG_Present_Time}
    w_{\rm DE} = -\frac{6 \pm 4\sqrt{3}\sqrt{1 + \Omega_0 - 2\Omega_{\Xi0}}}{6 - 3\Omega_0}.
\end{equation}

Inserting $w_{\rm DE} = -1$, consistent with the equation of state of DE in $\Lambda$CDM~\cite{Planck_2020}, and $\Omega_0 \sim 0.3$, $\Omega_{\Xi0} \sim 0.64$ is obtained, which is consistent with the CRG$-$solution. 

Values of $w_{\rm DE}$ different from $-$1 are consistent with several pieces of evidence anyway (e.g.,~\cite{Amendola_and_Tsujikawa_2015,Copeland_2006,Frusciante_and_Perenon_2020,Wen_2018,Capozziello_2006,Gerardi_2019}). 
Generally, the observational constraints on the equation of state of the effective DE can depend on the adopted model, even if model-independent reconstructions exist~\cite{Gerardi_2019}. For CRG, the parametrisation~\cite{Chevallier_and_Polarski_2001,Linder_2003} \textcolor{black}{(Section  4.3  in~\cite{Sanna_CRG_2023})}:
\begin{equation}
    \label{eq:w_DE_z_Parametrization}
    w_{\rm DE}(z) = w_0 + \frac{w_az}{1 + z}.
\end{equation}
can be assumed for $w_{\rm DE}$. At the present time ($z = 0$), Equation~\eqref{eq:w_DE_z_Parametrization} only depends on $w_0$, allowing a wide range of DE models, either with $w_{\rm DE} < -1$ (phantom models) or with $w_{\rm DE} \geq -1$~\cite{Bean_and_Melchiorri_2002,Sola_and_Stefancic_2005,Copeland_2006,Amendola_and_Tsujikawa_2015,Frusciante_and_Perenon_2020}. The parameter $w_0$ is constrained to be approximately in the range of $w_0 \in [-1.18,-0.85]$ by measurements of the baryonic acoustic oscillation (BAO), \textcolor{black}{SNe Ia}, and CMB~\cite{Hazra_2015,Wen_2018}. This limit on $w_0$ translates into a limit on $\Omega_{\Xi0}$, in agreement with the values of $\Omega_{\Xi0}$ estimated from the \textcolor{black}{SNe Ia} data.

\section{Discussion and Conclusions}
\label{sec:Discussion_And_Conclusions}

In this work, the main analyses and results of the theory of modified gravity RG were summarised. RG demonstrated the ability to model the dynamics of disk and elliptical galaxies with sensible mass and structural parameters (stellar mass-to-light ratios, disk scale heights, and orbital anisotropy parameters) and with RG parameters consistent among the different galaxies, suggesting their universality. Preliminary encouraging results were obtained at the scale of galaxy clusters and a covariant extension of the theory might look promising, since it seems to properly describe the accelerated expansion of the Universe, to retrieve the MOND acceleration scale $a_0$ at a galaxy scale, and to explain both the DM and DE sectors with a single scalar field.

Several further studies could be performed to complete the tests of this gravity theory. On a galaxy scale, two issues presented by the theory require additional investigation. They are (1) the prediction of a RAR with a too large intrinsic scatter and correlations between its residuals from Equation~\eqref{eq:fittedRAR} and some galaxy properties, and (2) the tension presented by the vacuum permittivity $\epsilon_0$, which might indicate its nonuniversality.

\textcolor{black}{Matsakos and Diaferio~\cite{Matsakos_and_Diaferio_RG_2016} showed that RG might be able to reproduce the BTFR and the MDAR. Instead, the} RAR built with RG presents some strong correlations, at more than 5$\sigma$, with some radially dependent galaxy properties \textcolor{black}{(the radius, the stellar surface density profile, and the gas fraction profile)}. However, the RAR built from DMS data also shows some, albeit weaker, correlations with some galaxy properties, and a further investigation is needed to understand whether the correlations presented by the RG RAR are partially driven by the data correlations. The correlations presented by the DMS data are apparently at odds with the claimed uncorrelations observed in the RAR derived from SPARC data~\cite{Lelli_SPARC_2016b}, suggesting a difference between the two samples. Moreover, the question of the RAR is even more puzzling, since the DMS is not the only sample where a correlation between the RAR residuals and some galaxy properties has been observed. Di Paolo et al.~\cite{Di_Paolo_2019} found a correlation between the RAR residuals and the galaxy radius from the accurate mass profiles of 36 dwarf disk spirals and 72 LSB galaxies. A better assessment of whether the result obtained in RG for the RAR depends on the theory itself or on the chosen galaxy sample can be performed by reproducing in RG the RAR of SPARC disk galaxies~\cite{Lelli_SPARC_2016b}, which, differently from DMS galaxies, are nearly edge-on and thus, their measured rotation curves, from which the RAR is derived, are much more accurate. Moreover, the SPARC sample counts much more galaxies than the DMS (153 vs. 30 galaxies), which would make this study much more statistically significant. For this study, the rotation curves of SPARC galaxies have to be modelled with RG, and the RAR has to be built from these models. The scatter and the residuals from Equation~\eqref{eq:fittedRAR} of the obtained RAR have to be compared with the results of Lelli et al.~\cite{Lelli_RAR_2017}. Another interesting study to be performed with the RAR would be to compute the RAR in RG separately for the groups of normal spirals and dwarf galaxies present in SPARC and in the sample considered by Di Paolo et al.~\cite{Di_Paolo_2019}. This study might shed light on the works of Santos-Santos et al.~\cite{Santos-Santos_2020} and Di Paolo et al.~\cite{Di_Paolo_2019}, which found that the RAR might differently behave in these two categories of galaxies, in addition to providing a further test for RG theory.

As said at the end of Section~\ref{sec:Elliptical_Galaxies_Results}, the tension on the vacuum permittivity $\epsilon_0$ might have different origins, either the approximate derivation of the unique $\epsilon_0$ in the DMS sample, a too simplistic dynamical model for the elliptical galaxies, an incorrect assumption for the functional form of the gravitational permittivity $\epsilon(\rho)$ (Equation~\eqref{eq:epsilon}), or, in the worst case, a fundamental issue with the RG theory. Generally, the study of elliptical galaxies is incomplete, both because the considered sample only counted three galaxies and because it only considered E0 ellipticals, with a nearly spherical shape. To better investigate the possible (non)universality of $\epsilon_0$, the performed study on the three E0 galaxies should be repeated removing the adopted approximations. In the new model, the interaction with the surrounding environment of the ellipticals should be considered, possibly resorting to $N$-body simulations. Additional E0 galaxies with extended kinematic profiles should be considered to enrich the galaxy sample in this new study. The E0 galaxies in the ePN.S survey~\cite{Pulsoni_PNe_Ellipticals_2018}, where extended kinematic profiles up to $\sim$13 $R_{\rm e}$ from the galaxy centres were measured thanks to the presence of planetary nebulae in the galaxies outskirts, should be suitable candidates. Other candidates ideal to test the viability of RG are the round elliptical galaxies in the samples of~\cite{Jimenez_2013,Durazo_2017,Durazo_2018}. Their velocity dispersion profiles present a flattening beyond a certain radius, where both this radius and the intensity of the velocity dispersion in this plateau are in agreement with MOND expectations. In the outer regions of spherical systems, where the density drops below $\rho_{\rm c}$, RG predicts a field proportional to $R^{-2}$, as in the Newtonian case (see Section~\ref{sec:RG}). Modelling in RG the velocity dispersion profiles of the round elliptical galaxies in~\cite{Jimenez_2013,Durazo_2017,Durazo_2018} would show whether they can be reproduced by a $\propto R^{-2}$ field in the galaxy outskirts or if they necessarily require a $\propto R^{-1}$ field as in MOND, which would represent an issue with the current formulation of RG.

After focussing on spherical elliptical galaxies alone, a more complete study should be performed considering elliptical galaxies with different ellipticities, to further test the capability of RG of modelling the dynamics of systems with different shapes with a unique set of RG parameters. This study would also permit one to investigate the positive correlation found by Deur~\cite{Deur_2014,Deur_2020} between the total mass-to-light ratios and the ellipticities of elliptical galaxies, already mentioned in Section~\ref{sec:Introduction}, which naturally arises within the RG context. Employing elliptical galaxies with kinematic profiles $\sim$10 times more extended than the data used by~\cite{Deur_2014,Deur_2020} might be crucial to validate or reject this correlation. All these studies might shed light on the correctness of Equation~\eqref{eq:epsilon} for the gravitational permittivity $\epsilon(\rho)$. 

A further test of the RG prediction that the flatter the system, the larger the mass discrepancy, if considered in Newtonian gravity, might be performed with the kinematic data of LSB galaxies (dwarf and dSph galaxies) and GCs, in addition to elliptical galaxies with different flatness degrees. Indeed, dwarf galaxies and GCs have similar baryonic masses but very different dynamical properties, the former being one of the darkest systems on a galaxy scale and the latter being nearly DM-free (Section~\ref{sec:Introduction}). \textcolor{black}{This last piece of evidence represented} a challenge for MOND (e.g.,~\cite{Baumgardt_GCs_2005,Jordi_GCs_2009,Baumgardt_GCs_2009,Sollima_and_Nipoti_GCs_2010,Ibata_GCs_2011a,Ibata_GCs_2011b,Frank_GCs_2012}). This feature of RG might be tested by modelling the rotation curves of a sample of dwarf galaxies, e.g., in the Milky Way halo~\cite{Salucci_2012}, in the LITTLE THINGS survey~\cite{Oh_2015}, in SPARC~\cite{Lelli_SPARC_2016b}, and in Di Paolo et al.~\cite{Di_Paolo_2019}, and the internal velocity dispersion of a sample of GCs located in the most external regions of the Milky Way (e.g.,~\cite{Baumgardt_GCs_2005,Jordi_GCs_2009,Baumgardt_GCs_2009,Sollima_and_Nipoti_GCs_2010,Ibata_GCs_2011a,Ibata_GCs_2011b,Frank_GCs_2012}), where the background density drops below $\rho_{\rm c}$. \textcolor{black}{If RG properly models the dynamics of these systems, it could explain why the dwarf galaxies are so ``dark'' if interpreted in Newtonian gravity, which can represent a challenge for $\Lambda$CDM~\cite{Mateo_Dwarfs_1998} (See Section~\ref{sec:Introduction}). Moreover, Matsakos and Diaferio~\cite{Matsakos_and_Diaferio_RG_2016} observed in a simplified framework that galaxies with a larger surface brightness in RG showed a smaller mass discrepancy, if interpreted in Newtonian gravity, than galaxies with a smaller surface brightness, which can naturally explain the difference between HSB and LSB galaxies and provide a solution for the cusp/core problem of $\Lambda$CDM (Section~\ref{sec:Introduction}).  An interesting study to test the viability of RG could be performed by modelling the dynamics of some dSph galaxies which present a high DM fraction, if interpreted in Newtonian gravity, but a not-so-squeezed shape, having an intrinsic ellipticity around 0.1--0.3~\cite{Walker_dSphs_2007,Salomon_dSphs_2015}.}

On the scale of galaxy clusters, the tests are incomplete, and a deeper investigation is required. The sample considered by Matsakos and Diaferio~\cite{Matsakos_and_Diaferio_RG_2016} is too small, only counting two galaxy clusters, and the analysis of the temperature profiles is approximated. For their study, Matsakos and Diaferio~\cite{Matsakos_and_Diaferio_RG_2016} considered the data from~\cite{Vikhlinin_Chandra_2006}, whose sample contains 11 other low-redshift and relaxed galaxy clusters. This analysis should be extended to these other galaxy clusters, by directly constraining the RG parameters from their temperature profiles to check their agreement with one another and with the results obtained on a galaxy scale. In a second step, the same analysis should be repeated by adopting a less approximate modelling by including the mass profiles of the individual galaxies present in the clusters, exploiting the results from $N$-body simulations.

In addition to the temperature profiles, RG should be tested on the dynamics of galaxy clusters. A study of this kind is already underway. Specifically, it should be verified whether RG is able to model the galaxy dispersion profiles of clusters of galaxies with sensible parameters. Possible samples for this analysis are CIRS~\cite{Rines_and_Diaferio_CIRS_2006} and HeCS~\cite{Rines_HeCS_2013}, which present the radial velocity dispersion profiles taken from the Sloan Digital Sky Survey and the ROSAT All-Sky Survey of 130 galaxy clusters in a redshift range of [0.0, 0.3] and with different dynamical properties, from nonmerging and relaxed to merging and dynamically active. This study allows us to investigate how the RG parameters are affected by the environment, either relaxed or interacting, extending to larger scales the possible future study of elliptical galaxies. 
 Also, the study of the BTFR extended to galaxy groups and clusters (Equation~\eqref{eq:BTFR_extended_galaxy_groups_and_clusters}) could be expanded, by considering how the uncertainties on the equation of state and on the entropy profiles of the gas and the possible local deviations from hydrostatic equilibrium due to X-ray gas flows might affect the values of the RG~parameters.

\textcolor{black}{Another important aspect of galaxy clusters that has not been investigated yet with RG and that should be studied to obtain a robust test for the viability of this theory, is the capability of RG to explain the hydrostatic mass bias, that is, the discrepancy between the masses of galaxy clusters measured with the gravitational lensing and the hot, X-ray-emitting gas methods~\cite{Gianfagna_2021}. It would also be interesting to explore whether the dependence of the RG gravitational field on the shape of the considered system might explain gravitational lensing effects.}

Also, CRG leaves room for additional investigations. The emergence of the MOND acceleration scale $a_0$ in the WFL of CRG should be further explored. Indeed, differently from $a_0$, $a_\Xi$ (Equation~\eqref{eq:acceleration_scale_CRG}) is not a constant but depends on the mass density of the source, even if this dependence drops at large distances from the source, since $\rho_{\rm s} \ll \rho_{\rm bg}$. Future studies are required to test if the connection between $a_\Xi$ and $a_0$ subsists for a generic case besides a source with a specific density configuration.

The parameter $w_{\rm DE} = p_{\rm DE}/\rho_{\rm DE}$ of the equation of state of the effective DE found by Sanna et al.~\cite{Sanna_CRG_2023} in CRG appears to be dependent on the redshift, differently from $\Lambda$CDM where $w_{\rm DE}$ is a constant equal to $-$1. However, $w_{\rm DE}$ tends to $-$1 at the present epoch, in agreement with the observations. The $w_{\rm DE}$ parameter in CRG could be constrained from the measurements of the expansion rate of the Universe performed by the Dark Energy Survey (DES) (\url{https://www.darkenergysurvey.org}, accessed on 4 April 2024)~\cite{DES_2005}, which has observed thousands of supernovae since 31 August 2013. Moreover, the Euclid mission (\url{https://sci.esa.int/web/euclid}, accessed on 4 April 2024)~\cite{Euclid_2022}, launched on 1 July 2023, is expected to measure the variation in the cosmic acceleration to an accuracy better than the 10\% level, which would be crucial to disentangle CRG from $\Lambda$CDM and other DE models~\cite{Amendola_Euclid_2018}. Future observations of the large-scale structure of the Universe and the evolution of DE will be crucial to constrain the value of $w_0$ in Equation~\eqref{eq:w_DE_z_Parametrization} and consequently, the value of $\Xi$ (see Section~\ref{sec:Covariant_RG}).

The Lagrangian density of CRG, derived from the CRG action (Equation~\eqref{eq:ST_action_CRG}), should be constrained from the power spectrum of the temperature anisotropies of the CMB and the power spectrum of the matter density perturbations (e.g.,~\cite{Huterer_2015,Noller_and_Nicola_2019}), for example, by comparing the CRG predictions with the latest measurements from the Planck satellite~\cite{Planck_2020_Overview_CP}. By estimating the cosmological parameters from different low-redshift galaxy surveys, such as the Kilo-Degree Survey (KIDS)~(\url{https://kids.strw.leidenuniv.nl}, accessed on 4 April 2024)~\cite{KIDS_2013}, \textls[-25]{Canada-France-Hawaii Telescope Legacy Survey (CFHTLS)~(\url{https://www.cfht.hawaii.edu/Science/CFHLS/}, accessed on 4 April 2024)~\cite{Gwyn_CFHTLS_2012}}, and DES, and from the CMB power spectrum, we can also assess whether RG can solve the tensions observed in the cosmological parameters in the $\Lambda$CDM model. 

Another essential test for CRG would consist in analysing the evolution of the density perturbations, at least in the linear regime, and how the scalar field and its perturbations would impact on the large-scale structure formation, evolution, and distribution \mbox{(e.g.,~\cite{di_Porto_and_Amendola_2008,Bueno_Sanchez_and_Perivolaropoulos_2011,Pace_2014,Kofinas_and_Lima_2017})}. A possible way to accomplish this task, would be to modify the publicly available \textcolor{black}{RAMSES code~\cite{Teyssier_RAMSES_2002}}, to run $N$-body simulations in the CRG framework. The results from these $N$-body simulations could be compared with the data from the DES survey, which probed the formation of structures with weak gravitational lensing and galaxy clusters, and the distribution of galaxies with the two-point correlation function. Further constraints could be given by the measurements taken with the Dark Energy Spectroscopic Instrument (DESI)~(\url{https://www.desi.lbl.gov}, accessed on 4 April 2024)~\cite{DESI_2023}, which started to observe in 2019, with Euclid, and with upcoming experiments such as the Square Kilometer Array (SKA)~(\url{https://www.skao.int/}, accessed on 4 April 2024)~\cite{SKA_2017}.

\textcolor{black}{To conclude, a fundamental test to disentangle RG theory from GR and other modified theories of gravity can be provided by a modelisation of the gravitational waves, first detected on 14 September 2015 by the LIGO experiment~\cite{LIGO_Collaboration_2016}. Indeed, some differences between GR and modified theories of gravity can be better stigmatised in the context of linearised gravitation in this era of the recently started gravitational wave astronomy, as stressed and anticipated by~\cite{Corda_GWs_2009}.}




\vspace{6pt}

\funding{This research received no external funding.  
}

\institutionalreview{Not applicable.}

\informedconsent{Not applicable.}

\dataavailability{No new data were created or analysed in this study. Data sharing is not applicable to this article.} 



\acknowledgments {I sincerely thank the three referees whose suggestions and corrections improved my review. Part of this work is the result of my Ph.D. activity at the Physics Department of the University of Turin (2017--2021). I sincerely thank all the people that helped me during my Ph.D., in particular, Antonaldo Diaferio, my Ph.D. supervisor, Titos Matsakos, who formulated the RG theory together with Antonaldo Diaferio, Garry Angus, who collaborated in the study of RG in DMS galaxies, and Andrea Sanna, who formulated CRG, together with Antonaldo Diaferio, and Titos Matsakos. This work has been supported by the Spoke 1 ``FutureHPC \& Big-Data'' of the ICSC-Centro Nazionale di Ricerca in High Performance Computing, Big
Data and Quantum Computing-and hosting entity, funded by European Union-NextGenerationEU.
}

\conflictsofinterest{The author declares no conflict of interest.} 


\abbreviations{Abbreviations}{
The following abbreviations are used in this manuscript:\\

\noindent 
\begin{tabular}{@{}ll}
BTFR & baryonic Tully--Fisher relation\\
CFHTLS & Canada-France-Hawaii Telescope Legacy Survey\\
CMB & cosmic microwave background\\
CRG & covariant refracted gravity\\
DE & dark energy\\
DES & Dark Energy Survey\\
DESI & Dark Energy Spectroscopic Instrument\\
DM & dark matter\\
DMS & DiskMass Survey\\
dSph & dwarf spheroidal\\
FLRW & Friedmann--Lema$\hat{i}$tre--Robertson--Walker\\
GCs & globular clusters\\
GR & general relativity\\
HSB & high surface brightness\\
ICM & intracluster medium\\
KIDS & Kilo-Degree Survey
\end{tabular}

\noindent 
\begin{tabular}{@{}ll}
$\Lambda$CDM & ~$\Lambda$ cold dark matter\\
LSB & ~low surface brightness\\
MDAR & ~mass discrepancy--acceleration relation\\
MGE & ~multi-Gaussian expansion\\
MOND & ~modified Newtonian dynamics\\
RAR & ~radial acceleration relation\\
RG & ~refracted gravity\\
RMS & ~root-mean-square\\
SKA & ~Square Kilometer Array\\
SMBH & ~supermassive black hole\\
SNe Ia & ~Ia supernovae\\
SPARC & ~Spitzer Photometry and Accurate Rotation Curves\\
SPS & ~stellar population synthesis\\
TeVeS & ~tensor vector scalar gravity\\
WFL & ~weak field limit\\
\end{tabular}
}

\begin{adjustwidth}{-\extralength}{0cm}
\reftitle{References}

\PublishersNote{}
\end{adjustwidth}
\end{document}